\PassOptionsToPackage{svgnames,x11names,table}{xcolor}
\PassOptionsToPackage{T1}{fontenc}
\documentclass[sigconf]{acmart}

\usepackage{acmart-taps}
\usepackage{listings}
\usepackage{url}
   \hypersetup{colorlinks=true, allcolors=DarkBlue}

\usepackage{tikz}
   \usetikzlibrary{shadows.blur}
\usepackage{dcolumn}
   \newcolumntype{d}[1]{D{.}{.}{#1}}
\usepackage{tabularx}
\usepackage{microtype}
\usepackage{afterpage}
\usepackage{graphicx}
\usepackage{makecell}
\usepackage{float}
   \usetikzlibrary{patterns}
\usepackage{pgfplots}
   \pgfplotsset{compat=1.18}
\usepackage{pgfplotstable}
   \usepgfplotslibrary{statistics}

\usepackage{fancyvrb}

\usepackage{enumitem}
\usepackage{booktabs}

\AtBeginDocument{%
  }

\copyrightyear{2026}
\acmYear{2026}
\setcopyright{cc}
\setcctype{by}
\acmConference[CHI '26]{Proceedings of the 2026 CHI Conference on Human Factors in Computing Systems}{April 13--17, 2026}{Barcelona, Spain}
\acmBooktitle{Proceedings of the 2026 CHI Conference on Human Factors in Computing Systems (CHI '26), April 13--17, 2026, Barcelona, Spain}
\acmPrice{}
\acmDOI{10.1145/3772318.3793707}
\acmISBN{979-8-4007-2278-3/2026/04}

\begin{document}

\title[User Privacy and Developer Willingness to Reduce Fingerprinting Risks]{Uncovering Relationships between Android Developers, User Privacy, and Developer Willingness to Reduce Fingerprinting Risks}


\author{Alex Berke}
\email{aberke@google.com}
\orcid{0000-0001-5996-0557}
\affiliation{%
  \institution{Google}
  \city{Cambridge, MA}
  \country{USA}}

\author{G\"{u}liz Seray Tuncay}
\email{gulizseray@google.com}
\orcid{0009-0003-5472-141X}
\affiliation{%
  \institution{Google}
  \city{San Francisco, CA}
  \country{USA}}

\author{Michael Specter}
\email{mikespecter@google.com}
\orcid{0000-0002-7662-4042}
\affiliation{%
  \institution{Google}
  \city{Atlanta, GA}
  \country{USA}}

\author{Mihai Christodorescu}
\email{christodorescu@google.com}
\orcid{0000-0001-5808-8015}
\affiliation{%
  \institution{Google}
  \city{Mountain View, CA}
  \country{USA}}

\renewcommand{\shortauthors}{Berke et al.}

\begin{abstract}

The major mobile platforms, Android and iOS, have introduced changes that restrict user tracking to improve user privacy, yet apps continue to covertly track users via device fingerprinting. 
We study the opportunity to improve this dynamic with a case study on mobile fingerprinting that evaluates developers’ perceptions of how well platforms protect user privacy and how developers perceive platform privacy interventions. Specifically, we study developers’ willingness to make changes to protect users from fingerprinting and how developers consider trade-offs between user privacy and developer effort.  We do this via a survey of 246 Android developers, presented with a hypothetical Android change that protects users from fingerprinting at the cost of additional developer effort.

 We find developers overwhelmingly (89\%) support this change, even when they anticipate significant effort, yet prefer the change be optional versus required. Surprisingly, developers who use fingerprinting are six times 
 more likely to support the change, despite being most impacted by it. We also find developers are most concerned about compliance and enforcement. In addition, our results show that while most rank iOS above Android for protecting user privacy, this distinction significantly reduces among developers very familiar with fingerprinting. Thus there is an important opportunity for platforms and developers to collaboratively build privacy protections, and we present actionable ways platforms can facilitate this.
\end{abstract}



\maketitle

\section{Introduction}
\label{sec:intro}

The major mobile platforms, Apple and Google, have made changes to improve privacy by allowing users to opt out of tracking via advertising IDs~\cite{apple,android}. However, mobile apps continue to covertly track users via “device fingerprinting”~\cite{AAFP-paper, Torres2018, kollnig-popets-2022}.  Fingerprinting is achieved by collecting device-specific attributes via APIs, and using their combination to identify and track users’ devices. 
Unlike tracking via advertising IDs, fingerprinting can occur without user notice and without an opt-out mechanism, presenting privacy risks that circumvent the control of both users and platforms. Apple’s policies disallow tracking users via fingerprinting~\cite{apple} and both Apple and Google have introduced changes to make this misuse of platform APIs more difficult~\cite{Apple-required-reasons-API,Android-SDK-runtime}. While these platform changes are important improvements, the fact that developers continue to circumvent them highlights the critical role developers play in improving or degrading user privacy. Yet the dynamics that guide whether developers adopt or evade policies designed to protect user privacy remain underexplored, which we address through this work.

More specifically, we conduct a fingerprinting study with Android developers. We build on recent works that leverage the open source nature of the Android ecosystem to demonstrate the pervasive privacy risk of fingerprinting~\cite{AAFP-paper, Torres2018} (the closed nature of the iOS ecosystem has made large scale iOS evaluations less available~\cite{kollnig-popets-2022}). 
Through this study we evaluate how developers’ relationships to fingerprinting impact their perceptions of how well platforms (Android and iOS) protect user privacy, how developers perceive platform changes that protect users from fingerprinting, and how developers consider trade-offs between user privacy and developer effort.
We do this by surveying 246 knowledgeable Android developers about a hypothetical platform change, termed “API Usage purposes”, described as a way to improve user privacy by protecting users from fingerprinting. 
The change comes at the potential cost of additional developer effort — developers must declare their reasons for using APIs that could be abused for fingerprinting in a manifest file~\cite{AndroidManifest.xml-file} — allowing us to study an effort-privacy trade-off. 
This change is similar to Apple’s “Required Reason API”, but no similar mechanism is present in Android at the time of this study. Thus this presents us with the opportunity to study developers’ responses to a viable platform intervention by surveying Android developers about this change to the Android ecosystem.

We use our survey to study the following research questions:
\begin{description}
    \item[RQ1] How do developers trade off developer effort versus user privacy?
    \item[RQ2] When introducing privacy-enhancing changes, do developers prefer platforms to mandate requirements or incentivize optional adoption?
    \item[RQ3] What are developers’ main concerns when platforms introduce privacy changes that require developers’ participation?
    \item[RQ4] How do developers comparatively perceive Android versus iOS’s protection of user privacy, and how does familiarity with fingerprinting impact this perception?
\end{description}

By answering these questions (Section~\ref{sec:findings}) we contribute to our understanding of developers’ willingness to adopt platform privacy enhancements. Furthermore, we quantify the effort-privacy tradeoff for developers, in contrast to previous work that only observed this tradeoff qualitatively~\cite{li-reddit-2021}.

We find that the overwhelming majority of developers (89\%) supported the intervention, including those who said it would require significant effort. More developers supported an optional implementation model, where developers are incentivized to implement the change, versus a required model. Yet still more developers supported the required model (41.5\%) versus no change at all (10.6\%).

Further analysis of this support yields both expected and unexpected results. 
As expected, developers who perceived a higher level of effort to implement the change were less supportive, and developers who perceived a more positive impact on user privacy were more supportive, demonstrating an effort-privacy trade-off. Unexpectedly, developers who use fingerprinting, and would be most impacted by the change, were significantly more supportive of the change. This suggests that the developers most needed to expend effort to reduce user tracking may be (unexpectedly) willing collaborators.

In addition, our analysis of open-ended comments reveals consistent concerns among developers. These include user experience, developer compliance, and platform enforcement, which were topics the survey did not mention.

Finally, our results show that most developers ranked iOS above Android for protecting user privacy.  Yet this was less often the case for developers very familiar with fingerprinting. This further indicates that fingerprinting can undermine platform privacy protections and developers’ perceptions of it, which we further discuss in Section~\ref{sec:discussion}.

\section{Related Work}
\label{sec:related}

To the best of our knowledge, we are the first to survey how developers’ relationships to fingerprinting impacts their approach to user privacy.

\subsection{Fingerprinting and User Protections}

A large body of research studies how websites, mobile applications, and software development kits (SDKs) fingerprint users~\cite{eckersley-2010,laperdrix-survey-2020,AAFP-paper,wenjia-2016}, and the resulting privacy risks~\cite{us-PETS-2025}.
For example, researchers have shown that fingerprinting is pervasive across the web and mobile app platforms, estimating that more than 25\% of the top 10K websites use fingerprinting~\cite{Iqbal2021} and more than 19\% of the top 30K Android apps do~\cite{Torres2018}. Other researchers have shown how fingerprinting risks vary across demographic groups, finding risks are higher for lower-income and older US user groups due to the types of devices they use~\cite{us-PETS-2025}.

Furthermore, fingerprinting often circumvents user choice to opt out of tracking~\cite{Papadogiannakis-2021} by taking advantage of APIs which exist to improve software functionality, violating users’ privacy defined by Contextual Integrity~\cite{nissembaum-contextual-integrity}, which considers the appropriate flow of data under users’ expectations.
For example, researchers have shown how websites often bypass GDPR protections and use fingerprinting even when users decline cookie consent banners~\cite{Papadogiannakis-2021}. They have also shown how fingerprinting scripts can be used to restore tracking cookies that users deleted and that this strategy is commonly used across the web, again bypassing users’ control~\cite{Fouad2022}.

There are also a number of studies that focus on automatic detection and prevention of fingerprinting behavior~\cite{TaintDroid, AndroPROTECT, Iqbal2021}, often treating the developers who use fingerprinting as adversaries. In contrast, we study an opportunity to shift these developers towards the adoption of platform protections.

In mobile apps, many APIs useful for fingerprinting are protected by permissions, where users must grant the app access~\cite{tuncay2024android, mayrhofer2024android}. App developers declare such permissions in a configuration file, which is the \lstinline!AndroidManifest.xml! file for Android~\cite{AndroidManifest.xml-file}.  Previous surveys have studied how both developers and users engage with permissions~\cite{iOS-tan-2014,Tahaei-2023,Felt-2012,Shen-2021}. Developers often request excessive permissions due to misunderstanding their scope or the needs of third-party libraries~\cite{Tahaei-2023}. Similarly, users tend to grant permissions without fully understanding them~\cite{Felt-2012,Shen-2021}. Prior work has shown these issues can be reduced for Android by changes within the Google Play Console that ``nudge'' developers to use fewer permissions~\cite{Peddinti-2019}. We also study an opportunity for a platform to impact developers’ use of permissions, via adding friction rather than nudging.  

\subsection{Developer Privacy Perceptions}

Prior empirical software engineering research has also surveyed developers about their relationships to user privacy~\cite{tahaei-2019}. 
These studies often conclude that developers care about user privacy, but their practices may contradict this sentiment, either because they are unaware of how third-party ads and analytics tooling collect user data~\cite{balebako-2014}, or because they see themselves as unable or not responsible to address such privacy risks~\cite{mhaidli2019we}.
When addressing how privacy can be improved, many studies have focused on the role of company culture and communication within teams and lack of awareness of privacy practices~\cite{hortsmann-2024, iwaya-2023, tahaei-privacy-champions-2021, balebako-2014}, advocating for greater adoption of privacy-by-design strategies~\cite{prybylo-2024, hadar-2018}.
When more specifically surveying app developers about user privacy and app permissions, developers have surfaced concerns about how the use of unnecessary permissions can break user trust~\cite{Tahaei-2023}. 
Our survey builds on these results by also focusing on app permission systems and similarly finding contradictions between developers’ desire to improve user privacy and their apps’ actual practices (i.e. their use of fingerprinting).
We further contribute to this literature by measuring developers’ preferences for a platform privacy enhancement to reduce fingerprinting, and how their perceptions of privacy benefits impact their preferences.

Other research has studied developer privacy perceptions by analyzing their discussions on online forums~\cite{green-shilton-platform-privacies,tahaei-SO-2020,li-reddit-2021}. For example, a forum analysis suggested that Android and iOS developers had different interpretations of privacy~\cite{green-shilton-platform-privacies}. Additionally, a case study of the Reddit Android developer forum, \lstinline!/r/androiddev!, suggested that developers often find new privacy-enhancing restrictions by the platform cause considerable cost yet fail to generate any compelling benefit for developers~\cite{li-reddit-2021}. The authors suggested Android complement restriction-based approaches with optional approaches that provide nudges~\cite{halpern-nudges-2015}, via rewards, for developers to make privacy-enhancing changes. In this work we address these qualitative findings and suggestions with quantitative analyses. We quantitatively test how Android developers consider trade-offs between their estimated effort to implement new privacy-enhancing platform changes (i.e., cost) versus impact on user privacy (benefit). Furthermore, we directly query their preference for optional, reward-based changes, versus required change.

\section{Materials and Methods}

To answer our research questions defined in \autoref{sec:intro}, we developed a survey (\autoref{sec:survey}), recruited Android developer participants (\autoref{sec:participants}), and analyzed the survey results using a mixed-methods approach (\autoref{sec:analysis}).

\subsection{Android Developer Survey}
\label{sec:survey}

\subsubsection{Survey Overview}
The survey introduced a hypothetical change to the Android platform, called ``API Usage Purposes,'' designed to protect users from unwanted fingerprinting and improve user privacy. It also queried developers’ familiarity with fingerprinting and whether they use fingerprinting in their apps or SDKs.

We limited our survey to an Android change with Android developers because Apple already has the similar “Required Reason API” and part of our goal is to study developers’ responses to a new intervention.

After describing API Usage Purposes, our survey solicited concerns, and asked participants how the change would impact user privacy and developer effort.  It also asked participants whether they thought Android should make this change as either a requirement, optional, or not at all. We used the responses to study how Android developers trade off developer effort versus user privacy (RQ1), whether developers prefer platforms introduce optional versus required changes to improve user privacy (RQ2), and to identify developers’ main concerns when platforms introduce privacy-enhancing changes that require developers’ participation (RQ3).

Our survey also gauged participants’ sentiments on how well Android and iOS protect user privacy. We used these responses to study RQ4.

\subsubsection{Survey Instrument}

\begin{figure*}
   \centering
  \tikz \node[fill=white,draw=gray,ultra thin] at (0,0) {\includegraphics[width=0.75\linewidth]{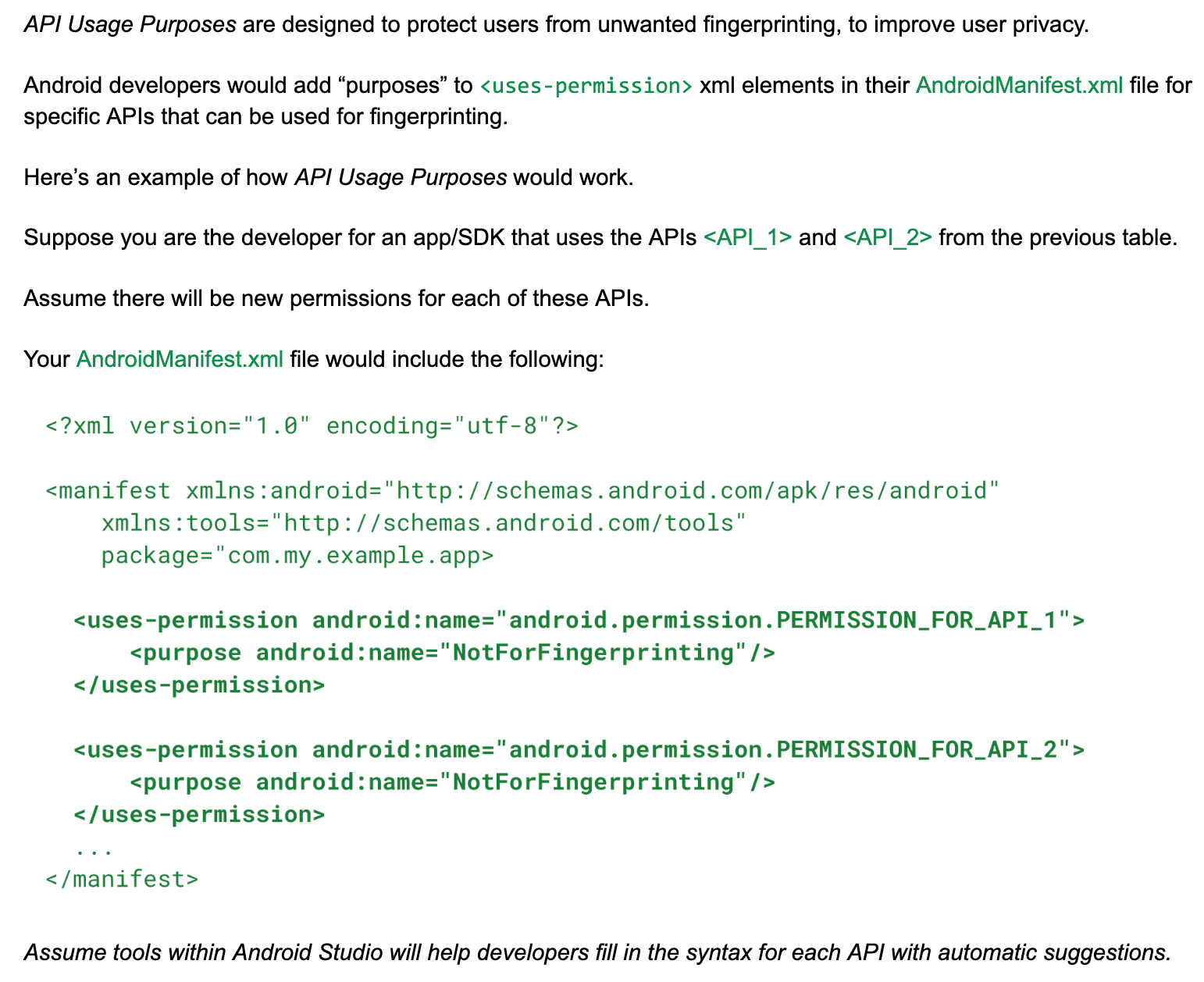}};
   \caption{Screenshot from the developer survey (directly before Q16), which explains how developers would implement the hypothetical change with an example \lstinline!AndroidManifest.xml! code snippet. \textsf{$<$API\_1$>$} and \textsf{$<$API\_2$>$} serve as example APIs that would be impacted by the change.}
   \Description{The image is a screenshot of a document or webpage describing "API Usage Purposes" in the context of Android development, specifically related to user privacy and fingerprinting.
The text is laid out in several paragraphs, followed by an XML code snippet.
Here's a breakdown of the content:
Top Section (Textual Description):
Paragraph 1: "API Usage Purposes are designed to protect users from unwanted fingerprinting, to improve user privacy."
Paragraph 2: "Android developers would add "purposes" to <uses-permission> xml elements in their AndroidManifest.xml file for specific APIs that can be used for fingerprinting."
Paragraph 3: "Here's an example of how API Usage Purposes would work."
Paragraph 4: "Suppose you are the developer for an app/SDK that uses the APIs <API_1> and <API_2> from the previous table."
Paragraph 5: "Assume there will be new permissions for each of these APIs."
Paragraph 6: "Your AndroidManifest.xml file would include the following:"
Middle Section (XML Code Snippet):
The XML code is formatted with indentation and green text (typical for code display).
<?xml version="1.0" encoding="utf-8"?>
<manifest xmlns:android="http://schemas.android.com/apk/res/android"
    xmlns:tools="http://schemas.android.com/tools"
    package="com.my.example.app">
    <uses-permission android:name="android.permission.PERMISSION_FOR_API_1">
        <purpose android:name="NotForFingerprinting" />
    </uses-permission>
    <uses-permission android:name="android.permission.PERMISSION_FOR_API_2">
        <purpose android:name="NotForFingerprinting" />
    </uses-permission>
    ...
</manifest>
Xml
Bottom Section (Textual Description):
Paragraph 7 (italicized): "Assume tools within Android Studio will help developers fill in the syntax for each API with automatic suggestions."
The overall purpose of the image is to illustrate how Android developers might declare the "purpose" of using certain APIs within their app's manifest file, specifically to clarify that the API is "NotForFingerprinting," as a measure to enhance user privacy.}
   \label{fig:survey_screenshot}
\end{figure*}

The survey was administered via Qualtrics. Details on the survey, and all question text, are provided in the Appendix (\ref{appendix:survey_sample_details}).

The survey first asked for participants’ informed consent (Q0) and confirmed participants were Android software developers working on either an Android app or software development kit (SDK) (Q1).  The survey then included a screening question to assess participants’ knowledge of the \lstinline!AndroidManifest.xml! file and permissions (Q2). Participants who answered incorrectly or “I don’t know” were filtered out of the sample used in analysis. 
The next set of questions (Q3-11) asked participants about how their app or SDK was categorized with Google Play and demographic questions, including age, gender, country, years of experience as a professional developer and developer team size, which were used in related app developer studies~\cite{mhaidli2019we, balebako-2014, Linares-Vasquez-2015, Linares-Vasquez-2017, Harjot-2022, tahaei2023stuck}. An attention check was also included to improve sample quality.

Before mentioning fingerprinting or “API Usage Purposes”, the survey asked how much participants agreed with the following statements: “Android protects user privacy” and “Apple protects user privacy” on a 1-10 scale (Q12).  

The survey then described device fingerprinting and asked if participants were already familiar with fingerprinting (Q13). On a following page the survey then asked whether their app or SDK fingerprints users (Q14). To reduce potential response bias, this question reminded participants that we would keep their answers confidential and not attempt to reconnect their responses with their app or SDK. 
The survey then presented “API Usage Purposes” as a hypothetical change to Android, designed to protect users from unwanted fingerprinting, to improve user privacy. It explained that with this change, Android developers would need to declare purposes for specific APIs that can be used for fingerprinting in their \lstinline!AndroidManifest.xml! file  (Figure~\ref{fig:survey_screenshot}).

Next, the survey asked participants how much they agreed with the statement that “Android protects user privacy” (with the same 1–10 scale as before), given the assumption that Android required API Usage Purposes (Q16). This repeated question (Q12) was used to measure a potential change in response before versus after API Usage Purposes.

The survey then solicited participants' concerns for API Usage Purposes via open ended comments (Q17a) and asked participants how the change would impact developer effort and user privacy (Q18a-b), with response options on a 5-point Likert scale.

Finally, participants were asked to consider two ways Android could implement the change: an optional model, where apps can receive a user-facing privacy badge and higher rank in the Google Play store, and a required model, where API calls fail when API Usage Purposes are not provided for the impacted APIs. Participants were then asked whether Android should implement the change with either the optional or required model, or not at all (Q19).

\subsection{Ethical considerations}
The authors’ institution did not require an IRB or ethics approval process for this study. However, we took the following steps to respect participants’ privacy. 
First, we made sure participants completed our informed consent form before allowing them to proceed to the survey, which informed them that their data may be used in a research publication and would be anonymized. We then only retained survey responses from participants who completed the entire study. Furthermore, we assigned random participant IDs and deleted any PII and other data that could be used to re-identify participants, including the app or SDK they worked on. 

All participants who completed the survey were compensated, whether or not they passed the attention check or screening.

\subsection{Participants}
\label{sec:participants}

We recruited English-speaking Android developer participants from two panels that are managed by a third-party vendor contracted by our company.
The first is a panel of mobile app and web developers who were recruited via banner ads placed on https://web.dev and https://developer.chrome.com, which host documentation for software developers. These developers were offered \$2.50 USD for completing the survey. 
The second is a panel of Android developers. They had opted-in to share their email addresses for research and marketing outreach when creating their Google Play accounts, and were later recruited for the panel via their Play account email addresses. These developers were offered \$5 for completing the survey (with the higher amount going to developers with already verified identities). Both panel recruitment processes included a pre-screening to ensure participants were professional developers in English speaking countries. We further used our survey screening to help ensure pre-screened participants were Android developers working on an app or SDK.
We note the use of targeted web ads and outreach via Google Play accounts has similarly been used to recruit developer participants for prior research studies~\cite{Harjot-2022,tahaei2023stuck}.
All participants completed our survey in June 2025 and were paid whether or not they were screened out. On average the survey took participants 7 minutes to complete.

A total of $498$ participants began the survey after providing informed consent. Our screening then excluded $124$ who were not Android developers working on an Android app or SDK, $117$ who did not pass the knowledge assessment for the \lstinline!AndroidManifest.xml! file and permissions, and $11$ who failed the attention check. The resulting data sample we analyzed includes the $N=246$ participants who passed all screening.

Details about the sample are in \autoref{appendix:survey_sample_details} and our anonymized survey dataset can be accessed upon request for research purposes. Of the $246$ participants, $214$ ($87\%$) were male, $23$ ($9.3\%$) female, and the remainder answered ``Other'' or ``Prefer not to answer.'' We note this gender imbalance is consistent with developer samples from prior work, where women represent $1$--$26\%$~\cite{Harjot-2022,Tahaei-2023,balebako-2014,mhaidli2019we}. When grouped by age, $58$ ($23.6\%$) were $18$--$34$ years old, $158$ ($64.2\%$) $35$--$54$ years old, and 27 ($11\%$) $55$ years or older.  The majority of participants had at least three years of professional Android developer experience and worked on teams of fewer than five developers. When grouped by country, the largest groups ($35.8\%$) were from the US, then the UK ($17.1\%$), India ($13.4\%$), Canada ($6.9\%$), and Germany ($6.5\%$).

\begin{figure*}
   \centering

    \includegraphics{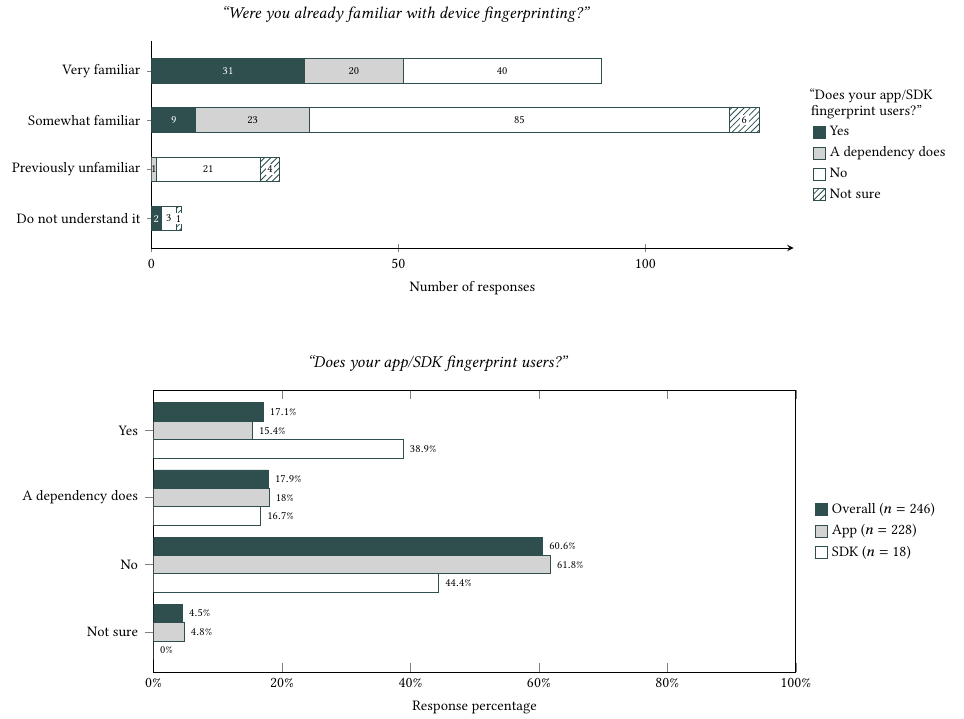}
   \caption{Participants’ relationships with fingerprinting.}
   \Description{
Here is a description of the image for visually impaired readers.

**Figure Title:** Fig. 2. Participants’ relationship with fingerprinting.

The image consists of two horizontal bar charts presenting survey data about device fingerprinting.

**Top Chart: "Were you already familiar with device fingerprinting?"**
This is a stacked bar chart displaying the "Number of responses" on the x-axis (ranging from 0 to roughly 125). The y-axis categories represent the participants' level of familiarity. Each bar is segmented to show the answer to a secondary question: "Does your app/SDK fingerprint users?" (Segments: Yes, A dependency does, No, Not sure).

* **Very familiar:** This group has a significant number of "Yes" responses.
    * Yes: 31
    * A dependency does: 20
    * No: 40
* **Somewhat familiar:** This group has the highest total number of responses, predominantly answering "No."
    * Yes: 9
    * A dependency does: 23
    * No: 85
    * Not sure: 6
* **Previously unfamiliar:**
    * Yes: 1
    * A dependency does: 21
    * No: 4
* **Do not understand it:**
    * Yes: 2
    * A dependency does: 3
    * No: 1
**Bottom Chart: "Does your app/SDK fingerprint users?"**
This is a clustered bar chart displaying the percentage of responses on the x-axis (0\% to 100\%). The y-axis categories are the answers to the question. The bars are clustered by respondent type: SDK developers (n=18), App developers (n=228), and Overall (n=246).

* **Yes:**
    * SDK: 38.9
    * App: 15.4%
    * Overall: 17.1%
* **A dependency does:**
    * SDK: 16.7%
    * App: 18%
    * Overall: 17.9%
* **No:**
    * SDK: 44.4%
    * App: 61.8
    * Overall: 60.6%
* **I’m not sure:**
    * SDK: 0%
    * App: 4.8%
    * Overall: 4.5%
   }
   \label{fig:sample_rel_to_fingerprinting}
\end{figure*}

\autoref{fig:sample_rel_to_fingerprinting} 
summarizes participants’ relationships to fingerprinting, with comparisons across app versus SDK developers (see also Appendix Tables~\ref{tab:sample_rel_to_fingerprinting}-\ref{tab:fingerprinting_familiarity_and_use}). 
228 (92.7\%) of the participants said they primarily work on an app while 18 (7.3\%) said they primarily work on an SDK (Q3).
When asked whether they were already familiar with device fingerprinting (Q13), 91 (37\%) answered ``Very familiar'', 123 (50\%) answered ``Somewhat familiar'', 26 (10.6\%) answered they were not previously familiar, but now understood it, and 6 (2.4\%) answered that even after the explanation, they did not understand it. When asked whether their app/SDK fingerprints users (Q14), the majority, 49 (60.6\%) answered ``No'', while 44 (17.9\%) answered ``Yes'', 44 (17.9\%) answered ``No directly, but a dependency does'', and 11 (4.5\%) answered ``I’m not sure''. 
Figure~\ref{fig:sample_rel_to_fingerprinting} shows how the majority of those who said they are ``very familiar'' with fingerprinting likely did so because their app/SDK uses fingerprinting, while the majority who said they do not use fingerprinting answered that they were only ``somewhat familiar'' or ``previously unfamiliar'' with fingerprinting.
It also shows that the percentage of SDK developers who use fingerprinting directly (38.9\%) is more than twice that of app developers (15.4\%), and more app developers said their app does fingerprinting via a dependency (i.e. SDK) versus directly (18\% versus 15.4\%). This is consistent with prior research that used program analysis to measure fingerprinting in Android apps and highlighted SDKs as the primary source of fingerprinting~\cite{AAFP-paper}. 
(For further context on developers’ use of fingerprinting and their app/SDK category, see Appendix Tables~\ref{tab:sample_app_categories}-\ref{tab:sample_sdk_categories}).
We used these responses in the following analyses to further study participants’ relationships to fingerprinting.

\subsection{Study Analysis}
\label{sec:analysis}

\subsubsection{Quantitative Analysis}
\label{sec:quant_analysis}

We hypothesized that developers who use fingerprinting in their apps and SDKs, and who would therefore be most impacted by a change like "API Usage Purposes", would be least supportive of this change.

We used logistic regression to test this hypothesis and study how developers trade off effort and user privacy (RQ1). We mapped responses to the survey question about the level of effort required to implement API Usage Purposes (Q18a), which was asked with a 5-level Likert scale, to a 1~(very little effort) to 5~(very significant effort) scale. We similarly mapped responses to the question about impact on user privacy (Q18b) to a 1~(very negative impact) to 5~(very positive impact) scale.
We also created a binary variable indicating whether the app or SDK they work on fingerprints users (Q14), mapping responses for either directly or via a dependency to 1, 0 otherwise. We used these as independent variables in the regression model. The dependent variable was whether or not they answered ``Yes'' when asked if Android should implement API Usage Purposes (Q19). This includes yes to either the required or optional model. 
For robustness, we repeated this analysis after excluding responses that answered ``Yes'' to the optional model, in order to restrict the analysis to testing support for a required change.

To analyze how developers comparatively perceived Android and iOS's protection of user privacy, we mapped their levels of agreement to the statements ``[Apple/Android] protects user privacy'' to whether they ranked Android above iOS, iOS above Android, or equivalently (Q12).
We used logistic regression to study how their familiarity with fingerprinting (Q13) impacted this outcome by using whether they said they were “very familiar with fingerprinting” (n=91) as a binary, independent variable, and whether they ranked iOS above Android as the dependent variable. 

We then measured the change in their response to ``Android protects user privacy'' at the end of the survey, after they were asked to assume API Usage Purposes were implemented (comparing Q16 to Q12).
We created a binary variable indicating whether their level of agreement with this statement increased ($1$) or not ($0$), and used this as the dependent variable in another logistic regression model which included whether they were ``very familiar with fingerprinting'' as an independent variable.

As an additional robustness check, we repeated all of the above regression analyses with variables to control for demographics (age and gender) as well as team size, years of professional developer experience, and whether the developer worked on an app versus SDK.

All analyses were conducted using Python and the pandas~\cite{the_pandas_development_team_2025_15831829} and statsmodels~\cite{seabold2010statsmodels} libraries.

\subsubsection{Qualitative Analysis of Comments}
\label{section:comments_analysis}

To study RQ3, the survey asked participants to optionally provide their concerns for API Usage Purposes (Q17a), resulting in 184 written responses. To evaluate these responses the four authors independently, inductively open-coded all responses using thematic analysis~\cite{miles-huberman-saldana-2013}, identifying themes, with some written responses having one theme and others having multiple themes.
They then discussed and merged their codebooks and aligned on six themes, provided in Appendix Table~\ref{tab:thematic_analysis}.
Two authors then acted as coders and independently went through the responses again to label each with the top most relevant theme, and then measured inter-rater reliability (where the calculation excludes blank responses).
The resulting Cohen’s kappa coefficient is $0.74$ which is considered substantial agreement~\cite{8d20e0b8-89d8-3d65-bcf5-8c19d56ec4ab}.
The two authors resolved the remaining disagreements through discussions to produce the final labels presented in the findings (\autoref{section:comments_results}).

\section{Findings}
\label{sec:findings}

Here we describe our findings, including developers’ support for changes (Section \ref{section:support_for_change}), their concerns (Section \ref{section:comments_results}), and perceptions of privacy in iOS versus Android (Section \ref{section:apple_v_android_privacy}).

\subsection{Support for Changes to Improve User Privacy (RQ1 and RQ2)}
\label{section:support_for_change}

\begin{figure*}
   \centering
   
   \includegraphics{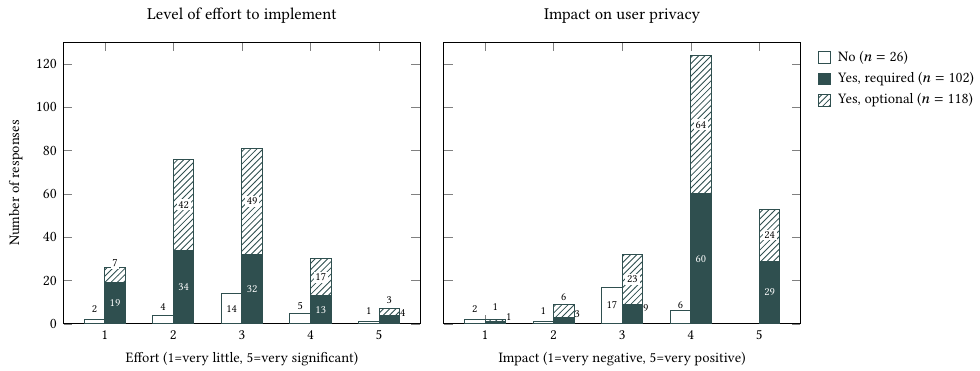}
   \caption{Responses to whether Android should implement the change by levels of perceived developer effort (left) and impact on user privacy (right).}
   \Description{The image displays two stacked bar charts side-by-side, sharing a common y-axis labeled "Responses" ranging from 0 to 120. A legend on the top right explains the fill patterns for the stacks:
White (empty) represents "No (n=26)"
Solid Black represents "Yes, required (n=102)"
Black diagonal stripes represents "Yes, optional (n=118)"
Left Chart: "Level of effort to implement (Very little to Very significant)"
This chart shows the perceived level of effort required for implementation, rated on a scale of 1 to 5.
Category 1 (Very little):
"No" responses: Very small, almost negligible.
"Yes, required" responses: Approximately 18.
"Yes, optional" responses: Approximately 8 (stacked on top of "Yes, required").
Total responses: Around 26.
Category 2:
"No" responses: Approximately 3.
"Yes, required" responses: Approximately 34.
"Yes, optional" responses: Approximately 40 (stacked on top of "Yes, required").
Total responses: Around 77.
Category 3:
"No" responses: Approximately 15.
"Yes, required" responses: Approximately 32.
"Yes, optional" responses: Approximately 34 (stacked on top of "Yes, required").
Total responses: Around 81.
Category 4:
"No" responses: Approximately 2.
"Yes, required" responses: Approximately 14.
"Yes, optional" responses: Approximately 14 (stacked on top of "Yes, required").
Total responses: Around 30.
Category 5 (Very significant):
"No" responses: Very small, almost negligible.
"Yes, required" responses: Approximately 4.
"Yes, optional" responses: Approximately 2 (stacked on top of "Yes, required").
Total responses: Around 6.
Right Chart: "Impact on user privacy (Very negative to Very positive)"
This chart shows the perceived impact on user privacy, rated on a scale of 1 to 5.
Category 1 (Very negative):
"No" responses: Approximately 1.
"Yes, required" responses: Very small, almost negligible.
"Yes, optional" responses: Very small, almost negligible.
Total responses: Around 1-2.
Category 2:
"No" responses: Approximately 9.
"Yes, required" responses: Approximately 3.
"Yes, optional" responses: Approximately 3 (stacked on top of "Yes, required").
Total responses: Around 15.
Category 3:
"No" responses: Approximately 18.
"Yes, required" responses: Approximately 8.
"Yes, optional" responses: Approximately 8 (stacked on top of "Yes, required").
Total responses: Around 34.
Category 4:
"No" responses: Approximately 2.
"Yes, required" responses: Approximately 60.
"Yes, optional" responses: Approximately 62 (stacked on top of "Yes, required").
Total responses: Around 124.
Category 5 (Very positive):
"No" responses: Very small, almost negligible.
"Yes, required" responses: Approximately 30.
"Yes, optional" responses: Approximately 23 (stacked on top of "Yes, required").
Total responses: Around 53.
Overall Interpretation:
Effort to implement: Most responses for "Yes, required" and "Yes, optional" fall into categories 2 and 3, suggesting a moderate level of effort is perceived. "No" responses are more evenly distributed but are generally very low in count for all categories.
Impact on user privacy: The "Yes, required" and "Yes, optional" categories show a strong peak at category 4, indicating that implementation (whether required or optional) is largely perceived to have a "Very positive" impact on user privacy (or at least a positive one). The "No" responses are more spread out, with a higher number in category 3, suggesting a more neutral or slightly positive view among those who did not implement. Categories 1 and 2 (very negative/negative impact) have very few responses across all types.}
   \label{fig:q_effort_privacy_responses}
\end{figure*}

\begin{figure*}
   \centering
   
   \includegraphics{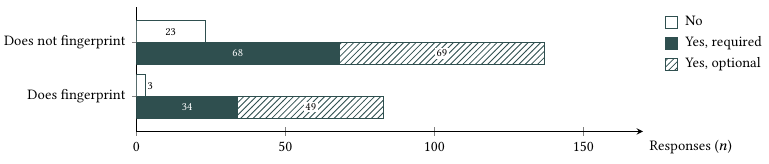}
   \caption{Responses to whether Android should implement the change by whether the developer said their app/SDK fingerprints users.}
   \Description{The image is a horizontal stacked bar chart showing the breakdown of "Responses" on the x-axis, ranging from 0 to 120. The y-axis has two categories: "Does not fingerprint" (top bar) and "Does fingerprint" (bottom bar).
A legend on the right side of the chart explains the different fill patterns for the stacked bars:
White (empty) represents "No"
Solid Black represents "Yes, required"
Black diagonal stripes represents "Yes, optional"
Detailed breakdown of each bar:
Does not fingerprint (Top Bar):
"No" (white): This segment extends from 0 to approximately 20 responses.
"Yes, required" (solid black): This segment starts from approximately 20 and extends to approximately 65 responses. The width of this segment is about 45 responses.
"Yes, optional" (black diagonal stripes): This segment starts from approximately 65 and extends to approximately 120 responses. The width of this segment is about 55 responses.
Total length of the "Does not fingerprint" bar: Approximately 120 responses.
Does fingerprint (Bottom Bar):
"No" (white): This segment is very small, extending from 0 to approximately 2 responses.
"Yes, required" (solid black): This segment starts from approximately 2 and extends to approximately 35 responses. The width of this segment is about 33 responses.
"Yes, optional" (black diagonal stripes): This segment starts from approximately 35 and extends to approximately 85 responses. The width of this segment is about 50 responses.
Total length of the "Does fingerprint" bar: Approximately 85 responses.
Summary of Observations:
For respondents whose app/SDK "Does not fingerprint", a large majority chose "Yes, optional" or "Yes, required" for something (presumably about implementing a feature or change related to fingerprinting, given context from other images).
For respondents whose app/SDK "Does fingerprint", also a large majority chose "Yes, optional" or "Yes, required" for something, though the total number of responses in this category is lower than "Does not fingerprint".
The number of "No" responses is very small for both "Does fingerprint" and "Does not fingerprint" categories, particularly for "Does fingerprint".}
   \label{fig:privacy_responses_by_fingerprinting}
\end{figure*}

Developers in our sample overwhelmingly supported the hypothetical API Usage Purposes. 
When asked whether Android should implement the change, $102$ ($41.5\%$) said ``Yes, with the required model,'' $118$ ($48.0\%$) said ``Yes, with the optional model,'' and $26$ ($10.6\%$) said ``No''. These results show more support for the optional versus required model, yet even so, nearly $4$ times as many participants supported a required change versus no change at all.
Furthermore, even developers who said the change would require moderate to significant effort to implement were more likely to support the change.

Most developers we surveyed anticipated the change would require very little to moderate effort, and most said the change would have a positive or very positive impact on user privacy. Results to these survey questions are shown in \autoref{fig:q_effort_privacy_responses} and Appendix Tables~\ref{tab:q_effort_responses} and~\ref{tab:q_privacy_responses}. 
Our regression analysis shows developers’ perceptions of the impact of the change on developer effort and user privacy significantly affects their support for the change.
Higher levels of perceived effort negatively impacted support ($OR = 0.46$; $p<0.01$) and more positive perceptions of the impact on user privacy positively impacted support ($OR = 3.651$; $p<0.001$).

Our regression analysis also tested whether developers whose apps or SDKs use fingerprinting were more or less likely to support API Usage Purposes. Contrary to our hypothesis, developers who reported using fingerprinting were more than $6$ times as likely to support the change ($OR = 6.480$; $p<0.05$, \autoref{fig:privacy_responses_by_fingerprinting}). 
Regression results are shown in \autoref{tab:support_regression_results}. 
As a robustness check, we repeated the analysis after filtering out participants who answered with support for the optional model, in order to directly compare the responses for ``Yes, with the required model'' to ``No, not at all.'' The regression results are directionally consistent, yet the effect sizes are even larger; developers whose apps/SDKs use fingerprinting were even more likely ($OR = 8.288$; $p<0.05$) to support the change (see Appendix \autoref{tab:support_regression_results_yes_required}).
We also performed a robustness check with variables indicating whether participants work on an app versus SDK and their demographics. The results are consistent and we did not find working on an app versus SDK had a significant impact (Appendix Table~\ref{tab:support_regression_results_w_demos}).

\begin{table}
   \centering
   \small
   \caption{Logistic regression model where the dependent variable is whether the developer said ``Yes'' to whether they think Android should implement the change. Note: $*$ $p<0.05$; $*$%
$*$ $p<0.01$; $*$%
$*$%
$*$ $p<0.001$.}
   \label{tab:support_regression_results}
   \begin{tabular}{ld{5}d{5}cd{5}}
      \toprule
      \textbf{Variable} & \multicolumn{1}{c}{\textbf{Coef.}} & \multicolumn{1}{c}{\textbf{OR}} & \textbf{95\% CI} & \multicolumn{1}{c}{\textbf{p-value}} \\
      \midrule
      Intercept & -0.716 & 0.489 & {[}0.061, 3.936{]} & 0.501 \\
      Effort level & -0.777** & 0.460 & {[}0.270, 0.782{]} & 0.004 \\
      Privacy impact & 1.295*** & 3.651 & {[}2.153, 6.189{]} & \textless{}0.001 \\
      Does fingerprint & 1.869** & 6.480 & {[}1.656, 25.354{]} & 0.007 \\
      \midrule
      N & 246 & & & \\
      Pseudo R-squared & 0.2514 & & & \\
     \bottomrule
   \end{tabular}
\end{table}

\subsection{Developer Concerns (RQ3)}
\label{section:comments_results}

\begin{figure*}
   \centering
   
   \includegraphics{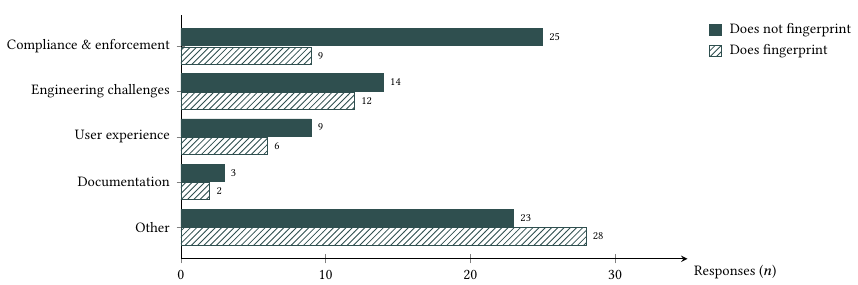}
   \caption{Top themes for developer concerns provided via open-ended responses}
   \Description{The image is a horizontal bar chart displaying "Responses (n)" on the x-axis, ranging from 0 to 50. The y-axis lists five categories from top to bottom: "Compliance & enforcement", "Engineering challenges", "User experience", "Documentation", and "Other". All bars are solid black on a white background.
Here's a breakdown of the length of each bar (number of responses):
Compliance & enforcement: The bar extends to approximately 35 responses.
Engineering challenges: The bar extends to approximately 25 responses.
User experience: The bar extends to approximately 16 responses.
Documentation: The bar extends to approximately 7 responses.
Other: The bar extends to approximately 48 responses.
Summary of Observations:
The category "Other" has the highest number of responses, indicating a significant portion of responses did not fit into the predefined categories.
"Compliance & enforcement" and "Engineering challenges" are the next most significant categories, suggesting these are major concerns or factors.
"User experience" and "Documentation" have fewer responses, indicating they are less frequently cited concerns compared to the top categories.}
   \label{fig:comments_themes_results}
\end{figure*}

\autoref{fig:comments_themes_results} summarizes the top themes that emerged from our inductive analysis of the $184$ open-ended responses (Section~\ref{section:comments_analysis}). 71 of these responses came from developers who use fingerprinting, either directly or via a dependency, which we indicate with an asterisk (*). 115 (46.7\%) participants did not list a concern about API Usage Purposes. This includes 62 blank responses and 53 responses that explicitly stated no concern  or positive support; some listed concerns about other parts of the Google/Android ecosystem. Examples include \textit{``Seems fine to me'' (P180)}, \textit{``This is a good idea'' (P20)}, and \textit{``No concerns. It's a good idea. It's rather Google that doesn't respect privacy'' (P33)}.

\subsubsection{Compliance and enforcement.} 
The most common theme regarded compliance and enforcement, however this theme was more common among developers who do not use fingerprinting (N=25 versus N*=9). Participants worried that developers would lie about whether they used APIs to fingerprint users, or would find other ways to evade the policy. For example, \textit{``Seems annoying and anyone can just lie, no?'' (P4)}.

In particular, many participants voiced concerns that the Google Play Store would not properly detect or validate the use, or lack of use, of API Usage Purposes. For example, \textit{``Will Google Play Store validates it's used correctly?'' (P51)}.

They were also concerned about lack of enforcement, or that enforcement would be arbitrary. For example, \textit{``Google Play’s enforcement of these XML tags will be arbitrary. I do not trust that having developer-declared usage will solve the fingerprinting problem'' (P16)} and \textit{``How it will be enforced. Every change like this ends up with a mess on the review process on the Play Store Console'' (P72)}.
Our survey did not mention compliance or enforcement, so the fact that so many participants independently voiced this concern should bring attention to it.

\subsubsection{Developer and engineering challenges.} 
26 participants highlighted potential developer and engineering challenges. These concerns were solicited before our survey asked participants to evaluate the level of effort required to implement API Usage Purposes and were more common for developers who use fingerprinting (N*=14 versus N=12), which might be expected given they would be more impacted by the changes. These concerns included how the change might limit the legitimate use of APIs, such as \textit{``More restriction could make it harder to use when there is a legitimate need for that information'' (P8)}.

Concerns also covered backwards compatibility, the increasing difficulty of building or testing an app, the potential for changes to break app functionality, and overall developer workload. Examples include \textit{``API usage makes it easier to build the App. Any restrictions will increase the QA efforts'' (P9*)}, \textit{``The potential issues with API and SDK calls not working properly'' (P101*)}, and \textit{``The additional work, risk of getting it wrong'' (P42)}.

\subsubsection{User experience.} 
Our survey did not mention user experience, leaving it undetermined whether users would be made aware of API Usage Purposes. Yet 15 participants still brought up concerns regarding the users’ perceptions or experiences. For example, \textit{``Could lead to issues where the user is concerned about something \dots that was never an issue for them to be concerned about the first place'' (P151)}, \textit{``Adding extra layers of information could lead to overwhelming the user'' (P113)} and \textit{``\dots vague or misleading descriptions, user fatigue from too many prompts\dots'' (P165*)}.

\subsubsection{Documentation.} 
5 participants also brought up concerns regarding clarity and documentation of how API Usage Purposes should be used. Most of these participants first listed support for the change. For example, \textit{``I do like very much that the permission can be declared for finger printing or not. Besides that, good documentation of how to implement this changes is paramount'' (P123*)} and \textit{``I don't have any. Just make sure it's well documented'' (P112)}.

\subsubsection{Other.} 
 51 participants voiced concerns that could not be grouped into consistent themes. Examples include \textit{``Unclear why it's needed'' (P97)}  or \textit{``I think it’s a nice improvement. I use AdMob, though, so I wonder how it would affect my advertising revenue'' (P88*)}.

\subsection{Perceptions of Privacy in iOS versus Android (RQ4)}
\label{section:apple_v_android_privacy}

\begin{figure}
   \centering
   
   \includegraphics{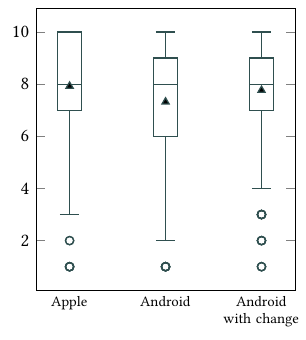}
   \caption{Boxplots showing the distribution of participants’ levels of agreement from 1 (strongly disagree) to 10 (strongly agree) for the statements ``[Apple/Android] protects user privacy.'' Lines indicate 25th, 50th (median) and 75th percentiles; triangles indicate means.}
   \Description{This image is a box-and-whisker plot displaying the distribution of data for three different categories: "Apple", "Android", and "Android with change".
The y-axis ranges from 0 to 10, likely representing some form of score or rating.
Detailed Description of Each Box Plot:
Apple (Leftmost Box Plot):
The box spans from approximately 7 to 10, indicating the interquartile range (IQR).
The median line within the box is around 8.
A black triangle marker, representing the mean, is positioned just below the median, around 7.8.
The upper whisker extends up to 10 (the top of the plot).
The lower whisker extends down to approximately 3.
There are two outlier data points represented by small circles below the lower whisker: one at approximately 2 and another at approximately 1.
Android (Middle Box Plot):
The box spans from approximately 6 to 9, indicating the IQR.
The median line within the box is at 8.
A black triangle marker, representing the mean, is positioned at approximately 7.4.
The upper whisker extends up to 10.
The lower whisker extends down to approximately 2.
There is one outlier data point represented by a small circle below the lower whisker, at approximately 1.
Android with change (Rightmost Box Plot):
The box spans from approximately 7 to 9, indicating the IQR.
The median line within the box is at 8.
A black triangle marker, representing the mean, is positioned just below the median, around 7.7.
The upper whisker extends up to 10.
The lower whisker extends down to approximately 4.
There are three outlier data points represented by small circles below the lower whisker: one at approximately 3, another at approximately 2, and a third at approximately 1.
General Observations:
All three categories have median scores around 8.
The mean scores (black triangles) are slightly lower than their respective medians, suggesting a slight skew towards lower values, especially for "Android".
"Apple" and "Android" have more outliers on the lower end compared to "Android with change".
The interquartile range (the box size) for "Apple" appears slightly larger than "Android" and "Android with change".
The range of data for "Apple" (excluding outliers) spans from ~3 to 10, for "Android" from ~2 to 10, and for "Android with change" from ~4 to 10.}
   \label{fig:android_v_apple_protect_user_privacy}
\vspace{-1cm}
\end{figure}

We evaluated participants' levels of agreement to the statements ``[Apple/Android] protects user privacy.''  The median level of agreement was the same ($median=8$) for Apple and Android (\autoref{fig:android_v_apple_protect_user_privacy}). However, the mean was higher for Apple ($mean=7.9$; $SD=2.1$) versus Android ($mean=7.3$; $SD=2.2$).

\begin{figure*}
   \centering
   
   \includegraphics{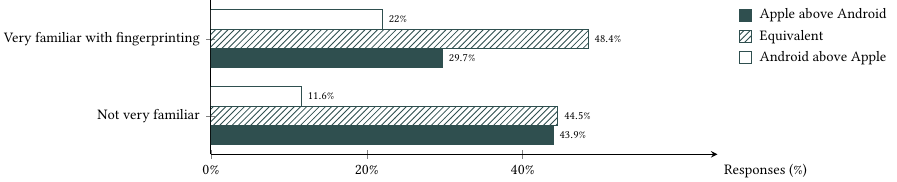}
   \caption{Participant-level comparisons for levels of agreement to whether ``[Apple/Android] protects user privacy,'' by whether or not they are very familiar with fingerprinting.}
   \Description{This image is a horizontal stacked bar chart showing the percentage of responses on the x-axis, ranging from 0 to 45. The y-axis has two categories: "Very familiar with fingerprinting" (top bar) and "Not very familiar" (bottom bar).
A legend in the top right corner defines the fill patterns for the stacked bars:
Solid Black: "Android above Apple"
Black diagonal stripes: "Equivalent"
White (empty): "Apple above Android"
Detailed breakdown of each bar:
Very familiar with fingerprinting (Top Bar):
"Android above Apple" (solid black): This segment extends from 0\% to approximately 22\% of responses.
"Equivalent" (black diagonal stripes): This segment starts from approximately 22\% and extends to approximately 45\% of responses. The width of this segment is about 23
"Apple above Android" (white): This segment starts from approximately 45\% and extends to approximately 30\% (it's a smaller segment to the left). The width of this segment is about 15\%.
Total length of the "Very familiar with fingerprinting" bar: Approximately 45\% (primarily composed of "Android above Apple" and "Equivalent"). The "Apple above Android" portion seems to be plotted incorrectly as a segment extending from the right of the black segment to the left, rather than stacked on top or extending to the right of the striped segment. However, interpreting it as a distinct percentage that adds to the total: 22\% + 23\% + 15\% would be 60\%. Given the scale, it's more likely that the total length of the bar represents 45\% and the white segment (Apple above Android) is a separate category that is also compared, but not necessarily part of the same stacked bar sum that reaches the total length of the 'Equivalent' portion. Let's re-interpret assuming it's a comparison within each familiarity group.
Re-interpretation: For "Very familiar with fingerprinting", ~22\% found Android to be above Apple, ~23\% found them equivalent, and ~15\% found Apple to be above Android. The total responses shown on the bar for Android above Apple + Equivalent is 45
Not very familiar (Bottom Bar):
"Android above Apple" (solid black): This segment extends from 0\% to approximately 10\% of responses.
"Equivalent" (black diagonal stripes): This segment starts from approximately 10\% and extends to approximately 45\% of responses. The width of this segment is about 35
"Apple above Android" (white): This segment starts from approximately 45\% and extends to approximately 42\% (a small segment). The width of this segment is about 3\%.
Total length of the "Not very familiar" bar: Approximately 45\% (primarily composed of "Android above Apple" and "Equivalent").
Re-interpretation: For "Not very familiar", ~10\% found Android to be above Apple, ~35\% found them equivalent, and ~3\% found Apple to be above Android.
Summary of Observations:
For both "Very familiar with fingerprinting" and "Not very familiar" groups, the "Equivalent" category accounts for the largest proportion of responses, suggesting a perception of similar performance or characteristics between Android and Apple.
Among those "Very familiar with fingerprinting", "Android above Apple" is more frequently reported (approx. 22\%) than "Apple above Android" (approx. 15\%).
Among those "Not very familiar", "Android above Apple" is reported at approx. 10\%, while "Apple above Android" is very low at approx. 3\%.
This suggests that generally, when a difference is perceived, Android is more often seen as "above Apple" than the reverse, especially among those who are very familiar. However, the most common perception is that they are equivalent.}
   \label{fig:android_v_apple_by_fingerprinting}
\end{figure*}

When making participant-level comparisons, we found that more participants ranked Apple above Android versus the other way around (see \autoref{fig:android_v_apple_protect_user_privacy} and Appendix \autoref{tab:android_v_apple_protect_user_privacy}) . However, we also found that developers’ familiarity with fingerprinting played a significant role in this ranking (\autoref{fig:android_v_apple_by_fingerprinting}).  Developers who said they were very familiar with fingerprinting were significantly less likely to rank Apple above Android with respect to user privacy ($OR = 0.540$; $p<0.05$). This finding holds when controlling for developer demographics and whether they work on an app versus SDK, which did not play a significant role. See \autoref{appendix:analysis_details} for details on the regression (Table~\ref{tab:Apple_v_Android_regression}) and a comparison in the distributions of participants’ agreement levels (Table~\ref{tab:android_v_apple_protect_user_privacy}). 

When participants were again asked about how ``Android protects user privacy'' after assuming API Usage Purposes were implemented, the median value did not change and the mean value slightly increased ($mean = 7.8$; $SD = 2.0$). We did not find that developers’ familiarity with fingerprinting played a significant role in this increase (see regression results in Appendix \autoref{tab:Android_privacy_change_regression}).

\section{Discussion}
\label{sec:discussion}

Our findings can provide insights for platforms as they make privacy-enhancing changes that affect both developers and end-users. We find developers’ relationships to fingerprinting impact their perceptions of how well platforms protect user privacy, and their willingness to participate in improving it.

\subsection{Developer Concerns about Enforcement and User Experience}

Our survey did not mention compliance and enforcement, or user experience, yet these themes emerged from the open ended comments left by developers. 

\paragraph{Compliance and enforcement.} Many developers voiced concerns that developers would lie about the true API purposes and that enforcement would be poorly handled by the platform. 
As platforms make changes impacting user trust, these concerns highlight how they should also consider developer trust. 

\paragraph{User experience.} Many developers were concerned that the change, if presented to users, would overwhelm them. Prior work has shown~\cite{alice-in-warningland} that over-prompting users with warnings can cause fatigue, effectively training users to ignore and click through otherwise important guardrails. It is unclear if, when, and how to present users with information like API Usage Purposes and thus it remains a question to address in future work.

\subsection{Developers Support Changes to Improve User Privacy Despite Additional Effort}

Our work addresses prior research that suggested Android provide incentive-based approaches when making privacy-related changes~\cite{li-reddit-2021}. Indeed, developers we surveyed preferred an optional, incentive-based change over a requirement. Our work also addresses the issue of cost for developers when Android introduces such changes~\cite{li-reddit-2021}. We found overwhelming support for ``API Usage Purposes,'' despite additional effort required by developers. 

Furthermore, contrary to our hypothesis (\autoref{sec:quant_analysis}), \textbf{we found developers who use fingerprinting were six times more likely to support the change}, despite potential impact to their business model. A possible explanation is that, while many developers care about user privacy, developers who use fingerprinting are much more aware of the negative impact this practice has on user privacy.
Yet they may also need to engage in fingerprinting to maintain a competitive edge,  since so many of their competitors use this highly effective user tracking method. While our survey did not address why developers use fingerprinting, previous work has demonstrated how fingerprinting provides data for precise ad targeting~\cite{Liu2025}, which is economically valuable~\cite{Wernerfelt2024}. This scenario presents a collective action problem that a platform change like API Usage Purposes could help resolve by impacting all apps at once. 
Regardless of why developers who use fingerprinting support a change to reduce fingerprinting, this surprising result is heartening – developers could support the adoption of such privacy enhancements, if the barriers we found (e.g. documentation, compliance, and enforcement) were overcome.

\subsection{Different Perspectives on Platform Privacy}
\label{sec:differences_on_platform_privacy}

On average, developers we surveyed ranked Apple slightly higher than Android for protecting user privacy, yet the rankings were very similar (the medians identical). This similarity is more reflective of a large-scale empirical analysis of Android and iOS apps that found widespread user tracking and privacy violations in both ecosystems~\cite{kollnig-popets-2022}.

The developers in our survey who indicated strong familiarity with fingerprinting were significantly less likely to rank Apple above Android for protecting user privacy. While further surveys are needed to explain this discrepancy, we provide some speculation. Apple has marketed itself as a more private platform~\cite{martin-murphy-2017,clinch-cnbc-2014} and Apple made a highly publicized~\cite{newman-forbes-2022} change in iOS 14, called ``App Tracking Transparency,'' which requires apps to directly request the user’s permission before tracking them with advertising IDs~\cite{apple-data-privacy-day-2021,apple-app-tracking-transp-2025}. While this may improve Apple’s privacy sentiment for many users and developers less familiar with fingerprinting, those who understand how fingerprinting bypasses user tracking controls may feel differently.  Regardless of the reason for this discrepancy, our findings suggest that Android could improve its comparative privacy sentiment by improving education and awareness around fingerprinting.

\subsection{Opportunity and Recommendations for Platform–Developer Collaboration on User Privacy}

Our results are consistent with prior work indicating app developers care about user privacy, even when their actions may be contradictory~\cite{mhaidli2019we,balebako-2014}. They also echo prior work that calls for increasing collaboration between developers and API designers rather than treating developers as enemies~\cite{chowdhury}.
Altogether, these results present an opportunity for platform-developer collaboration to improve user privacy.

In particular, we recommend platforms make changes to more directly address how fingerprinting undermines user privacy, which our survey showed most Android developers are ready to support.
These changes can be motivated by developers’ and platforms’ shared incentive of increasing user trust and perceived privacy – developers care about user trust~\cite{Tahaei-2023} and privacy~\cite{mhaidli2019we, balebako-2014}, and the major platforms consistently highlight the ways they improve user privacy~\cite{applePrivacy, AndroidPrivacy}.
These changes can be further supported by better educating users and developers about fingerprinting, as noted in Section~\ref{sec:differences_on_platform_privacy}. In terms of how to make the changes to address fingerprinting risks, API Usage Purposes can serve as an example, as well as provide an example for platform-developer collaboration on improving user privacy more broadly.

With any such changes we recommend platforms collaboratively incorporate developer feedback, such as the concerns solicited via this survey.
For example, platforms should help alleviate developer burden and engineering challenges by providing tools (e.g., through their IDEs) that simplify implementation.
Platforms should also support developers with clear documentation that includes how changes will impact user experiences. They can also support developers with well defined policies and mechanisms to detect and enforce violations that developers can trust.

\subsection{Limitations and Future work}

The scope of our survey is limited to Android developers and questions about fingerprinting protections are limited to a specific proposal. Future work should address generalizability. This includes testing whether  our findings on developer perceptions of user privacy extend to iOS developers. 
In particular, our survey did not ask if participants were familiar with iOS development or iOS’s Required Reason API~\cite{Apple-required-reasons-API}, which could have informed their responses about the platforms’ user privacy protections or ``API Usage Purposes''. 
To address these limitations, future work can survey both Android and iOS developers and also ask iOS developers about their actual experiences using the Required Reason API.
Another limitation is that our survey relied on self-reported data and described a hypothetical change at a high level with pseudocode. 
It could be valuable to validate participants’ reported relationships to fingerprinting by drawing information from their apps/SDKs and validate our findings in real-world development settings. (We did not connect participants to their apps/SDKs to protect their privacy and minimize their self-censorship.)

Potential biases in survey responses also present limitations. For example, social desirability bias could lead participants who use fingerprinting to say they do not. However, our result that developers who use fingerprinting are more likely to support a change to reduce fingerprinting holds despite this potential bias, which we expect would weaken this result. Even so, developers we surveyed opted in to participate, and this self-selection may present bias.

Finally, our survey raises important questions that remain unanswered. If developers are ready to adopt changes to protect users from fingerprinting, why do many of them use fingerprinting? It is possible that developers have little control over the fingerprinting done by SDKs that their apps use, or fear losing the competitive advantage that fingerprinting provides. If developers do not wish to lose the benefits of fingerprinting and user tracking (e.g. increased advertising revenue~\cite{Wernerfelt2024,Liu2025}), the platforms may need to guide a collective change in developer behavior. While “API Usage Purposes” may present one way, understanding the optimal mechanisms to effect such changes is the topic of further research.

\section{Conclusion}
Our study evaluated developers’ willingness to adopt a privacy-enhancing platform change via a survey with 246 knowledgeable Android developers. The survey introduced a hypothetical change to improve user privacy by reducing fingerprinting risks, with a potential cost to developers.

Our contributions include measuring an effort-privacy trade-off and finding developers are overwhelmingly supportive of such a privacy-enhancing change, even when they anticipate significant effort.
Furthermore, we found developers who use fingerprinting, and would be most impacted by the change, were surprisingly most likely to support it.
Our analysis of developers’ comments also highlights their concerns around platform policy enforcement, providing guidance for platforms as they consider such changes. Overall, this study reveals an important opportunity for platforms to improve user privacy with a collaborative model that addresses developer concerns and guides collective action.

\begin{acks}
We thank the Android developers who participated in our survey for their contributions. We also thank Mark Cwalinski and Matt Warner for helping motivate and develop our research questions, Richard Smith for helping recruit developers, and the anonymous reviewers for providing helpful feedback to clarify the work. We are thankful to Hamzeh Zawawy, René Mayrhofer, and Dave Kleidermacher for their feedback and support throughout this project. 
\end{acks}

\bibliographystyle{ACM-Reference-Format}

\colorlet{blue}{black}

\section*{Appendix}
\appendix

\section{Survey and sample details}
\label{appendix:survey_sample_details}
\textcolor{black}{
The text for the survey questions and responses are provided in full at the end of the Appendix.
A survey pilot was first conducted with 10 participants from the same participant pool as the main sample. We used the pilot to confirm the survey functioned as intended and then made no changes to the survey. Since the pilot survey and participant pool are the same as the expanded sample, pilot responses are included into the full analysis sample.
}

\autoref{tab:sample_demographics} shows the distribution of the sample's age and gender demographics and \autoref{tab:sample_country} shows the distribution of countries where the developers say they live.
In addition to these basic demographics, the survey also asked participants about their roles as developers, including whether they primarily work on an app or SDK (and if they work on multiple, to answer questions for the one they spend the most time on), their development team size, and their years of professional experience as an Android developer. Responses are shown in \autoref{tab:sample_dev_details}.

The survey also asked developers to provide which category either their app or SDK is listed as, depending on whether they said they primarily work on an app or SDK. Response options were from the lists of categories from the Google Play Console \cite{google-play-categories} and Google SDK Index \cite{sdk-categories}, respectively. Responses are shown in Tables~\ref{tab:sample_app_categories} and~\ref{tab:sample_sdk_categories}.

\begin{table}[t]
    \centering
    \small
    \caption{Sample gender and age demographics (N=246)}
    \label{tab:sample_demographics}
    \begin{tabular}{lrr}
        \toprule
        & \textbf{n} & \textbf{\%} \\
        \midrule
        \textbf{Gender} & & \\
        Man & 214 & 87.0 \\
        Woman & 23 & 9.3 \\
        Prefer not to answer & 7 & 2.8 \\
        Other & 2 & 0.8 \\ \addlinespace
        \textbf{Age} & & \\
        18 - 24 & 3 & 1.2 \\
        25 - 34 & 55 & 22.4 \\
        35 - 44 & 92 & 37.4 \\
        45 - 54 & 66 & 26.8 \\
        55 - 64 & 20 & 8.1 \\
        65 or older & 7 & 2.8 \\
        Prefer not to answer & 3 & 1.2 \\
        \bottomrule
    \end{tabular}
\end{table}

\begin{table}[t]
    \centering
    \small
    \caption{Sample developer details (N=246)}
    \label{tab:sample_dev_details}
    \begin{tabular}{lrr}
        \toprule
        & \textbf{n} & \textbf{\%} \\
        \midrule
        \textbf{App or SDK} & & \\
        App & 228 & 92.7 \\
        SDK & 18 & 7.3 \\ \addlinespace
        \textbf{Team size} & & \\
        1 & 91 & 37.0 \\
        2 - 4 & 90 & 36.6 \\
        5 - 9 & 41 & 16.7 \\
        10 - 24 & 17 & 6.9 \\
        25 - 49 & 4 & 1.6 \\
        50 or more & 3 & 1.2 \\ \addlinespace
        \textbf{Experience} & & \\
        Less than 1 yr & 11 & 4.5 \\
        1 to 2 yrs & 26 & 10.6 \\
        3 to 5 yrs & 49 & 19.9 \\
        6 to 9 yrs & 63 & 25.6 \\
        10 to 14 yrs & 69 & 28.0 \\
        15+ yrs & 28 & 11.4 \\
        \bottomrule
    \end{tabular}
\end{table}

\begin{table}[t]
    \centering
    \small
    \caption{Sample geographic distribution (N=246)}
    \label{tab:sample_country}
    \begin{tabular}{lrr}
        \toprule
        \textbf{Country} & \textbf{} & \textbf{\%} \\
        \midrule
        United States of America & 88 & 35.8 \\
        United Kingdom & 42 & 17.1 \\
        India & 33 & 13.4 \\
        Canada & 17 & 6.9 \\
        Germany & 16 & 6.5 \\
        Finland & 6 & 2.4 \\
        Netherlands & 6 & 2.4 \\
        Kenya & 4 & 1.6 \\
        Malaysia & 3 & 1.2 \\
        Italy & 3 & 1.2 \\
        Brazil & 2 & 0.8 \\
        Nigeria & 2 & 0.8 \\
        Sweden & 2 & 0.8 \\
        Ireland & 1 & 0.4 \\
        Serbia & 1 & 0.4 \\
        Indonesia & 1 & 0.4 \\
        Singapore & 1 & 0.4 \\
        Australia & 1 & 0.4 \\
        Trinidad and Tobago & 1 & 0.4 \\
        Spain & 1 & 0.4 \\
        \bottomrule
    \end{tabular}
\end{table}

\begin{table}
    \centering
    \small
    \caption{P\textcolor{black}{articipants' p}rimary app category listing \textcolor{black}{and fingerprinting use} (N=228).}
    \label{tab:sample_app_categories}
    \resizebox{\columnwidth}{!}{%
    \begin{tabular}{lrrrrrr}
    \toprule
    & \multicolumn{4}{c}{\textbf{(Q14) Does your app fingerprint users?}} & \multicolumn{2}{c}{\textbf{Total}} \\
    \cmidrule(lr){2-5} \cmidrule(lr){6-7}
    \textbf{(Q6a) App Category} & \textbf{Yes} & \makecell{\textbf{Dependency} \\ \textbf{does}} & \textbf{No} & \makecell{\textbf{Not} \\ \textbf{sure}} & \textbf{n} & \textbf{\%} \\
    \midrule
    Productivity & 4 & 4 & 20 & 0 & 28 & 12.3 \\
    Games & 5 & 5 & 16 & 1 & 27 & 11.8 \\
    Tools & 1 & 4 & 18 & 1 & 24 & 10.5 \\
    Education & 1 & 6 & 14 & 1 & 22 & 9.6 \\
    Business & 6 & 4 & 12 & 0 & 22 & 9.6 \\
    Entertainment & 2 & 0 & 13 & 2 & 17 & 7.5 \\
    Finance & 5 & 3 & 2 & 0 & 10 & 4.4 \\
    Travel and Local & 2 & 2 & 5 & 1 & 10 & 4.4 \\
    Lifestyle & 0 & 4 & 5 & 0 & 9 & 3.9 \\
    Shopping & 3 & 1 & 5 & 0 & 9 & 3.9 \\
    Health and Fitness & 0 & 0 & 8 & 0 & 8 & 3.5 \\
    Maps and Navigation & 0 & 1 & 4 & 1 & 6 & 2.6 \\
    Food and Drink & 1 & 1 & 3 & 0 & 5 & 2.2 \\
    Music and Audio & 0 & 2 & 2 & 1 & 5 & 2.2 \\
    Communications & 1 & 0 & 3 & 0 & 4 & 1.8 \\
    Sports & 0 & 1 & 1 & 1 & 3 & 1.3 \\
    Medical & 0 & 0 & 3 & 0 & 3 & 1.3 \\
    Books and Reference & 0 & 1 & 2 & 0 & 3 & 1.3 \\
    Auto and Vehicles & 0 & 1 & 1 & 0 & 2 & 0.9 \\
    Photography & 1 & 0 & 1 & 0 & 2 & 0.9 \\
    Weather & 0 & 0 & 2 & 0 & 2 & 0.9 \\
    House and Home & 1 & 0 & 0 & 1 & 2 & 0.9 \\
    Comics & 1 & 0 & 0 & 0 & 1 & 0.4 \\
    Personalization & 0 & 0 & 1 & 0 & 1 & 0.4 \\
    News and Magazines & 0 & 0 & 0 & 1 & 1 & 0.4 \\
    Social & 1 & 0 & 0 & 0 & 1 & 0.4 \\
    Video Players and Editors & 0 & 1 & 0 & 0 & 1 & 0.4 \\
    \bottomrule
    \end{tabular}
    }
\end{table}

\begin{table*}[ht]
    \centering
    \small
    \caption{\textcolor{black}{Participants' primary SDK category listing and fingerprinting use (N=18).}}
    \label{tab:sample_sdk_categories}
    \begin{tabular}{lrrrrrr}
    \toprule
    & \multicolumn{4}{c}{\textbf{Does your SDK fingerprint users?}} & \multicolumn{2}{c}{\textbf{Total}} \\
    \cmidrule(lr){2-5} \cmidrule(lr){6-7}
    \textbf{(Q6b) SDK Category} & \textbf{Yes} & \makecell{\textbf{Dependency} \\ \textbf{does}}& \textbf{No} & \makecell{\textbf{Not} \\ \textbf{sure}} & \textbf{n} & \textbf{\%} \\
    \midrule
    Data management & 3 & 0 & 3 & 0 & 6 & 33.3 \\
    Analytics & 0 & 0 & 3 & 0 & 3 & 16.7 \\
    Advertising and monetization & 1 & 0 & 1 & 0 & 2 & 11.1 \\
    Marketing and engagement & 1 & 1 & 0 & 0 & 2 & 11.1 \\
    Payments & 1 & 1 & 0 & 0 & 2 & 11.1 \\
    User support & 0 & 1 & 1 & 0 & 2 & 11.1 \\
    User authentication & 1 & 0 & 0 & 0 & 1 & 5.6 \\
    \bottomrule
    \end{tabular}
\end{table*}

\begin{table*}[ht]
\centering
\small
\caption{\textcolor{black}{Developer participants' r}elationship\textcolor{black}{s} to fingerprinting.}
\label{tab:sample_rel_to_fingerprinting}
\begin{tabular}{lcccccc}
\toprule
 & \multicolumn{2}{c}{\textbf{Total}} & \multicolumn{2}{c}{\textbf{SDK}} & \multicolumn{2}{c}{\textbf{App}} \\
\textbf{Question and response} & \textbf{n} & \textbf{\%} & \textbf{n} & \textbf{\%} & \textbf{n} & \textbf{\%} \\
\midrule
\multicolumn{7}{l}{\textbf{(Q13) ``Were you already familiar with device fingerprinting?''}} \\
Very familiar & 91 & 37 & 12 & 66.7 & 79 & 34.6 \\
Somewhat familiar & 123 & 50 & 5 & 27.8 & 118 & 51.8 \\
I was not familiar with device fingerprinting, but now I understand it & 26 & 10.6 & 1 & 5.6 & 25 & 11 \\
Even after the explanation, I do not understand device fingerprinting & 6 & 2.4 & 0 & 0 & 6 & 2.6 \\
\midrule
\multicolumn{7}{l}{\textbf{(Q14) ``Does your app/SDK fingerprint users''}} \\
Yes & 42 & 17.1 & 7 & 38.9 & 35 & 15.4 \\
Not directly, but a dependency does & 44 & 17.9 & 3 & 16.7 & 41 & 18 \\
No & 149 & 60.6 & 8 & 44.4 & 141 & 61.8 \\
I'm not sure & 11 & 4.5 & 0 & 0 & 11 & 4.8 \\
\bottomrule
\end{tabular}
\end{table*}

\begin{table*}[htbp]
\centering
\small
\caption{\textcolor{black}{Developers participants' familiarity with fingerprinting by their app/SDK use.}}
\label{tab:fingerprinting_familiarity_and_use}
\begin{tabular}{p{7.8cm}rrrrrr}
\toprule
 & \multicolumn{4}{c}{\textbf{(Q14) ``Does your app/SDK fingerprint users?''}} & \multicolumn{2}{c}{\textbf{Total}} \\
\cmidrule(lr){2-5} \cmidrule(lr){6-7}
\textbf{(Q13) Familiarity with fingerprinting} & Yes & Dependency does & No & Not sure & n & \% \\

\midrule
Even after the explanation, I do not understand device fingerprinting & 2 & 0 & 3 & 1 & 6 & 2.4 \\
I was not familiar with device fingerprinting, but now I understand it & 0 & 1 & 21 & 4 & 26 & 10.6 \\
Somewhat familiar & 9 & 23 & 85 & 6 & 123 & 50.0 \\
Very familiar & 31 & 20 & 40 & 0 & 91 & 37.0 \\
\bottomrule
\end{tabular}
\end{table*}

\section{Additional analysis details and robustness checks}
\label{appendix:analysis_details}

\begin{figure*}
    \centering
    \includegraphics[width=0.7\linewidth]{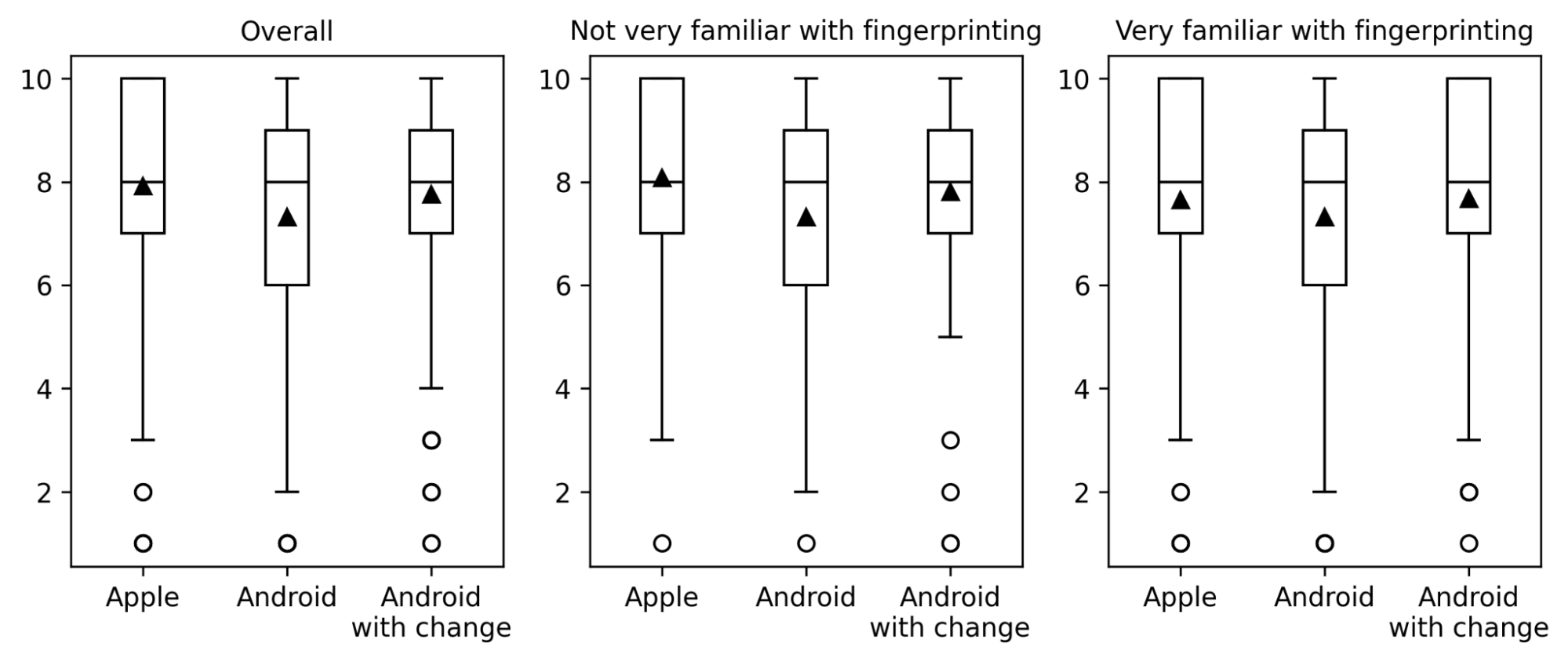}
    \caption{Distribution of levels of agreement for the statements “[Apple/Android] protects user privacy” from 1 (strongly disagree) to 10 (strongly agree) comparing participants overall (left), versus those who did not say they are very familiar with fingerprinting (middle) and those who did say they are very familiar with fingerprinting (right). Lines indicate the 25th, 50th (median) and 75th percentiles; triangles indicate the mean.}
    \label{fig:android_v_apple_protect_user_privacy_extended}
    \Description{The image is a scientific figure containing three separate box plot charts arranged horizontally. Each chart compares user perceptions of privacy protection for "Apple," "Android," and "Android with change." User agreement is rated on a vertical scale from 1 (strongly disagree) to 10 (strongly agree). The figure is divided into three populations: "Overall" (left chart), "Not very familiar with fingerprinting" (middle chart), and "Very familiar with fingerprinting" (right chart).

Each box plot illustrates the distribution of data: a central box shows the range from the 25th to the 75th percentile, a horizontal line inside the box marks the median (50th percentile), and a black triangle indicates the mean. Vertical lines, or "whiskers," extend from the box to show the range of the data, and individual circles below the whiskers represent outlier data points.

**Chart 1: Overall**
This chart shows the data for all participants.
* **Apple:** The median agreement is 8. The box ranges from 7 to 10. The mean is approximately 7.9. The whiskers extend from about 3 to 10, with outliers near 1 and 2.
* **Android:** The median is 7.5. The box ranges from 6 to 9. The mean is approximately 7.4. The whiskers extend from 2 to 10, with one outlier near 2.
* **Android with change:** The median is 8. The box ranges from 7 to 9. The mean is approximately 7.8. The whiskers extend from 4 to 10, with outliers near 2 and 3.

**Chart 2: Not very familiar with fingerprinting**
This chart shows data for participants who are not very familiar with browser/device fingerprinting. The results are very similar to the "Overall" chart.
* **Apple:** The median is 8. The box ranges from 7 to 10. The mean is approximately 8.0. The whiskers extend from 3 to 10, with an outlier near 1.
* **Android:** The median is 7. The box ranges from 6 to 9. The mean is approximately 7.3. The whiskers extend from 2 to 10, with one outlier near 2.
* **Android with change:** The median is 8. The box ranges from 7 to 9. The mean is approximately 7.8. The whiskers extend from 5 to 10, with outliers near 2 and 3.

**Chart 3: Very familiar with fingerprinting**
This chart shows data for participants who are very familiar with browser/device fingerprinting.
* **Apple:** The median is 8. The box ranges from 7 to 10. The mean is approximately 7.9. The whiskers extend from 3 to 10, with outliers near 1 and 2.
* **Android:** The median is 7. The box ranges from 6 to 8. The mean is approximately 7.2. The distribution appears slightly lower and tighter than in the other two charts. The whiskers extend from 2 to 10.
* **Android with change:** The median is 8. The box ranges from 7 to 10. The mean is approximately 7.9. The whiskers extend from 4 to 10, with an outlier near 2.

**Summary of Trends:**
Across all three groups, Apple's perceived privacy protection is consistently high, with a median rating of 8. Android's perceived protection is generally slightly lower than Apple's. The "Android with change" scenario consistently shows a higher median and mean rating than the standard "Android" scenario, suggesting the proposed change improves privacy perception. For users very familiar with fingerprinting, the perceived privacy of the standard Android system is slightly lower than for other groups.}
\end{figure*}

\autoref{tab:android_v_apple_by_fingerprinting},  \autoref{fig:android_v_apple_protect_user_privacy_extended},
\autoref{tab:android_v_apple_protect_user_privacy}, \autoref{tab:Apple_v_Android_regression}, \autoref{tab:Apple_v_Android_regression_w_demos} and \autoref{tab:Android_privacy_change_regression} provide additional data to support the findings presented in \autoref{section:apple_v_android_privacy}.
\autoref{tab:q_effort_responses}, \autoref{tab:q_privacy_responses} and 
\autoref{tab:support_vs_does_fingerprint} provide additional data to support the findings presented in \autoref{section:support_for_change}.

\begin{table*}
\small
\caption{Participant-level comparisons for levels of agreement to whether “[Apple/Android] protects user privacy” (Q12), by whether or not they are very familiar with fingerprinting (Q13). The data corresponds to \autoref{fig:android_v_apple_by_fingerprinting}.}
\label{tab:android_v_apple_by_fingerprinting}

\begin{tabular}{lrrrrrr}
\toprule
 & \multicolumn{2}{c}{\textbf{All}} & \multicolumn{2}{c}{\textbf{Very familiar}} & \multicolumn{2}{c}{\textbf{Not very familiar}} \\
 & \textbf{n} & \textbf{\%} & \textbf{n} & \textbf{\%} & \textbf{n} & \textbf{\%} \\
\midrule
Apple above Android   &   95    &   38.6                                &   27                  &   29.7   &   68   &   43.9   \\
Equivalent            &   113   &   45.9                                &   44                  &   48.4   &   69   &   44.5   \\
Android above Apple   &   38    &   15.4                                &   20                  &   22.0   &   18   &   11.6   \\
\bottomrule
\end{tabular}
\end{table*} 

\begin{table*}
\centering
\small
\caption{Distribution of levels of agreement for the statements “[Apple/Android] protects user privacy” (Q12) from 1 (strongly disagree) to 10 (strongly agree).}
\label{tab:android_v_apple_protect_user_privacy}
\setlength{\tabcolsep}{1pt} 
\begin{tabular}{l|c|c|c|c|c|c|c|c|c}
\toprule
\multicolumn{4}{c|}{\textbf{All (N=246)}} & \multicolumn{3}{c}{\textbf{\begin{tabular}[c]{@{}c@{}}Not very familiar with \\ fingerprinting (N=155)\end{tabular}}} & \multicolumn{3}{|c}{\textbf{\begin{tabular}[c]{@{}c@{}}Very familiar with \\ fingerprinting (N=91)\end{tabular}}} \\ 
\midrule
& \textbf{Apple} & \textbf{Android} & \makecell{\textbf{Android} \\ \textbf{with change}} & \textbf{Apple} & \textbf{Android} & \makecell{\textbf{Android} \\ \textbf{with change}} & \textbf{Apple} & \textbf{Android} & \makecell{\textbf{Android} \\ \textbf{with change}} \\
\midrule
Mean & 7.9 & 7.3 & 7.8 & 8.1 & 7.3 & 7.8 & 7.7 & 7.3 & 7.7 \\
Std. dev. & 2.1 & 2.2 & 2.0 & 1.8 & 2.0 & 1.8 & 2.4 & 2.5 & 2.4 \\
Min & 1.0 & 1.0 & 1.0 & 1.0 & 1.0 & 1.0 & 1.0 & 1.0 & 1.0 \\
25\% & 7.0 & 6.0 & 7.0 & 7.0 & 6.0 & 7.0 & 7.0 & 6.0 & 7.0 \\
50\% & 8.0 & 8.0 & 8.0 & 8.0 & 8.0 & 8.0 & 8.0 & 8.0 & 8.0 \\
75\% & 10.0 & 9.0 & 9.0 & 10.0 & 9.0 & 9.0 & 10.0 & 9.0 & 10.0 \\
Max & 10.0 & 10.0 & 10.0 & 10.0 & 10.0 & 10.0 & 10.0 & 10.0 & 10.0 \\
\bottomrule
\end{tabular}
\end{table*}

\begin{table*}
\centering
\small
\caption{Logistic regression model where the dependent variable is whether the developer ranked Apple above Android with respect to protecting user privacy.}
\label{tab:Apple_v_Android_regression}

\begin{tabular}{lrrcr}
\hline
\textbf{Independent variable} & \textbf{Coef.} & \textbf{OR} & \textbf{95\% CI for OR} & \textbf{p-value} \\
\hline
Intercept & -0.246 & 0.782 & {[}0.569, 1.073{]} & 0.128 \\
Very familiar with fingerprinting & -0.617* & 0.540 & {[}0.311, 0.936{]} & 0.028 \\
\hline
N & 246 & & & \\
Pseudo R-squared & 0.01512 & & & \\
\hline
\end{tabular}
\end{table*}

\begin{table*}
\centering
\small
\caption{Robustness check for the logistic regression model where the dependent variable is whether the developer ranked Apple above Android with respect to protecting user privacy.}
\label{tab:Apple_v_Android_regression_w_demos}
\begin{tabular}{lcccc}
\toprule
\textbf{Independent variable} & \textbf{Coef.} & \textbf{OR} & \textbf{95\% CI for OR} & \textbf{p-value} \\
\midrule
Intercept & -1.632 & 0.196 & {[}0.037, 1.028{]} & 0.054 \\
\textit{App or SDK (Ref: SDK)} & & & & \\
\quad App & 1.298 & 3.662 & {[}0.988, 13.582{]} & 0.052 \\
\textit{Team size (Ref: 1)} & & & & \\
\quad 2 - 9 & 0.605 & 1.831 & {[}0.996, 3.365{]} & 0.052 \\
\quad 10 or more & 0.288 & 1.334 & {[}0.485, 3.670{]} & 0.577 \\
\textit{Years experience (Ref: Less than 3 years)} & & & & \\
\quad 3 - 9 years & -0.087 & 0.916 & {[}0.409, 2.054{]} & 0.832 \\
\quad 10+ years & -0.272 & 0.762 & {[}0.314, 1.848{]} & 0.547 \\
\textit{Age (Ref: 18 - 34)} & & & & \\
\quad 35 - 54 & 0.04 & 1.040 & {[}0.523, 2.071{]} & 0.910 \\
\quad 55+ & -0.321 & 0.725 & {[}0.256, 2.057{]} & 0.546 \\
\textit{Gender (Ref: Non-male)} & & & & \\
\quad Male & -0.035 & 0.966 & {[}0.422, 2.212{]} & 0.935 \\
Very familiar with fingerprinting & -0.626* & 0.535 & {[}0.297, 0.962{]} & 0.037 \\
\midrule
N & 243 & & & \\
Pseudo R-squared & 0.04270 & & & \\
\bottomrule
\end{tabular}
\end{table*}

\vspace{5cm}

\begin{table*}
\centering
\small
\caption{Logistic regression model where the dependent variable is whether the developer increased their agreement with the statement that Android protects user privacy after supposing API Usage Purposes are implemented.}
\label{tab:Android_privacy_change_regression}
\begin{tabular}{lcccc}
\toprule
\textbf{Independent variable} & \textbf{Coef.} & \textbf{OR} & \textbf{95\% CI for OR} & \textbf{p-value} \\
\midrule
Intercept & -0.405* & 0.667 & {[}0.483, 0.919{]} & 0.013 \\
Very familiar with fingerprinting & -0.304 & 0.738 & {[}0.429, 1.269{]} & 0.272 \\
\midrule
N & 246 & & & \\
Pseudo R-squared & 0.003754 & & & \\
\bottomrule
\end{tabular}
\end{table*}

\begin{table*}
\centering
\small
\caption{Responses to survey question Q18a: ``What level of developer effort would be required to support API Usage Purposes?''}
\label{tab:q_effort_responses}
\begin{tabular}{llcccc}
\toprule
& & & \multicolumn{3}{c}{\textbf{By support response}} \\ 
\cmidrule(lr){4-6}
\textbf{Level} & & \textbf{All} & \textbf{No} & \textbf{Yes, required} & \textbf{Yes, optional} \\
\midrule
1 & Very little effort & 28 & 2 & 19 & 7 \\
2 & Little effort & 80 & 4 & 34 & 42 \\
3 & Moderate level of effort & 95 & 14 & 32 & 49 \\
4 & Significant level of effort & 35 & 5 & 13 & 17 \\
5 & Very significant level of effort & 8 & 1 & 4 & 3 \\
\bottomrule
\end{tabular}
\end{table*}

\begin{table*}
\centering
\small
\caption{Responses to survey question Q18b: “What impact, if any, would this change have on user privacy?”}
\label{tab:q_privacy_responses}
\begin{tabular}{llcccc}
\toprule
& & & \multicolumn{3}{c}{\textbf{By support response}} \\
\cmidrule(lr){4-6}
\textbf{Level} & & \textbf{All} & \textbf{No} & \textbf{Yes, required} & \textbf{Yes, optional} \\
\midrule
1 & It would have a large negative impact on user privacy & 4 & 2 & 1 & 1 \\
2 & It would have a small negative impact on user privacy & 10 & 1 & 3 & 6 \\
3 & It would have no impact on user privacy & 49 & 17 & 9 & 23 \\
4 & It would have a small positive impact on user privacy & 130 & 6 & 60 & 64 \\
5 & It would have a large positive impact on user privacy & 53 & 0 & 29 & 24 \\
\bottomrule
\end{tabular}
\end{table*}

\begin{table*}
\centering
\small
\caption{Responses to whether Android should implement “API Usage Purposes” (Q19), broken down by whether or not the developer works on an App/SDK that fingerprints users (Q14).}
\label{tab:support_vs_does_fingerprint}
\begin{tabular}{lcccccc}
\toprule
& \multicolumn{2}{c}{\textbf{Total}} & \multicolumn{2}{c}{\textbf{Does fingerprint}} & \multicolumn{2}{c}{\textbf{Does not fingerprint}} \\
\cmidrule(lr){2-3} \cmidrule(lr){4-5} \cmidrule(lr){6-7}
& \textbf{n} & \textbf{\%} & \textbf{n} & \textbf{\%} & \textbf{n} & \textbf{\%} \\
\midrule
No, not at all & 26 & 10.6 & 3 & 3.5 & 23 & 14.4 \\
Yes, optional & 118 & 48.0 & 49 & 57.0 & 69 & 43.1 \\
Yes, required & 102 & 41.5 & 34 & 39.5 & 68 & 42.5 \\
\bottomrule
\end{tabular}
\end{table*}

\autoref{tab:support_regression_results_yes_required} provides results for a robustness check for the logistic regression model presented in \autoref{tab:support_regression_results}. Responses where the participant answered ``Yes, with the optional model'' are filtered out of this analysis in order to compare the responses for ``Yes, with the required mode'' to ``No, not at all.''  
The results are directionally consistent with the main model (\autoref{tab:support_regression_results}), yet the effect sizes are even larger.
Developers whose apps/SDKs use fingerprinting are even more likely (OR = 8.3 versus OR = 6.5) to support API Usage Purposes with the required model.

\autoref{tab:support_regression_results_w_demos} provides results for another robustness check for the logistic regression model presented in \autoref{tab:support_regression_results}, where demographic details are included. \textcolor{black}{We did not find demographic variables were statistically significant, whether or not they were aggregated. For simplicity, we report the model with aggregated variables: g}ender variables are consolidated to male versus non-male (woman/prefer not to say/other), age variables are consolidated to 3 groups (18 - 34 years, 35 - 54 years, 55 or older), participants who did not answer age are excluded (n=3), team size is consolidated to 3 groups (1, 2 - 9, 10 or more) and years of experience is consolidated to 3 groups (Less than 3 years, 3 - 9 years, 10 or more years).

\begin{table*}
\centering
\small
\caption{Robustness check for logistic regression model where the dependent variable is whether the developer said “Yes, with the required model” to whether they think Android should implement the change. Responses for “Yes, with the optional model” are not included.}
\label{tab:support_regression_results_yes_required}
\begin{tabular}{lcccc}
\toprule
\textbf{Independent variable} & \textbf{Coef.} & \textbf{OR} & \textbf{95\% CI for OR} & \textbf{p-value} \\
\midrule
Intercept & -2.421 & 0.089 & {[}0.005, 1.591{]} & 0.100 \\
Effort level & -1.053** & 0.349 & {[}0.177, 0.687{]} & 0.002 \\
Privacy impact & 1.715*** & 5.556 & {[}2.651, 11.643{]} & <0.001 \\
Does fingerprinting & 2.115* & 8.288 & {[}1.594, 43.088{]} & 0.012 \\
\midrule
N & 128 & & & \\
Pseudo R-squared & 0.3648 & & & \\
\bottomrule
\end{tabular}
\end{table*}

\begin{table*}
\centering
\small
\caption{Robustness check for logistic regression model where the dependent variable is whether the developer said “Yes” to whether they think Android should implement the change, where additional demographic variables are included.}
\label{tab:support_regression_results_w_demos}
\begin{tabular}{lcccc}
\toprule
\textbf{Independent variable} & \textbf{Coef.} & \textbf{OR} & \textbf{95\% CI} & \textbf{p-value} \\
\midrule
Intercept & -2.102 & 0.122 & {[}0.003, 4.961{]} & 0.266 \\
\textit{App or SDK (Ref: SDK)} & & & & \\
\quad App & -1.391 & 0.249 & {[}0.022, 2.781{]} & 0.259 \\
\textit{Team size (Ref: 1)} & & & & \\
\quad 2 - 9 & -0.337 & 0.714 & {[}0.217, 2.348{]} & 0.579 \\
\quad 10 or more & -0.588 & 0.556 & {[}0.079, 3.917{]} & 0.555 \\
\textit{Experience (Ref: Under 3 years)} & & & & \\
\quad 3 - 9 years & -1.133 & 0.322 & {[}0.034, 3.063{]} & 0.324 \\
\quad 10+ years & -1.772 & 0.170 & {[}0.017, 1.724{]} & 0.134 \\
\textit{Age (Ref: 18 - 34)} & & & & \\
\quad 35 - 54 & -1.577 & 0.207 & {[}0.021, 2.076{]} & 0.180 \\
\quad 55+ & -2.537 & 0.079 & {[}0.006, 1.118{]} & 0.060 \\
\textit{Gender (Ref: Non-male)} & & & & \\
\quad Male & 1.534 & 4.639 & {[}0.949, 22.683{]} & 0.058 \\
Effort level & -0.834** & 0.434 & {[}0.245, 0.770{]} & 0.004 \\
Privacy impact & 1.654*** & 5.230 & {[}2.586, 10.580{]} & <0.001 \\
Does fingerprinting & 1.851* & 6.368 & {[}1.253, 32.358{]} & 0.026 \\
\midrule
N & 243 & & & \\
Pseudo R-squared & 0.3571 & & & \\
\bottomrule
\end{tabular}
\end{table*}

\begin{table*}
\centering
\small
\caption{Themes, descriptions and occurrences for open-ended responses regarding concerns for API Usage Purposes (Q17a).}
\label{tab:thematic_analysis}
\begin{tabularx}{\textwidth}{l p{2cm} X >{\itshape}X ccc}
\toprule
& \textbf{Theme} & \textbf{Description} & \textbf{Example from data} & \textbf{Total count} & \textbf{Does fingerprint} & \textbf{Does not fingerprint}\\
\midrule
1 & Compliance and enforcement & Developers may lie or evade the policy; Accuracy of detection; Google may not validate or enforce whether the purpose label is properly used; Enforcement process may be arbitrary. & "I don't see an enforcement mechanism." & 34 & 9 & 25\\ \addlinespace
2 & Documentation & Clear documentation of requirements and use. & "I do like very much that the permission can be declared for finger printing or not. Besides that, good documentation of how to implement this changes is paramount." & 5 & 2 & 3\\ \addlinespace
3 & User experience & Related to user experience; information shown to users; users misunderstanding the information; information/decision overload for users & "Adding extra layers of information could lead to overwhelming the user." & 15 & 6 & 9\\ \addlinespace
4 & Developer and engineering challenges & Limits legitimate uses of impacted APIs; Adds to developer workload; Breaks apps or harms backwards compatibility; etc. & "My major concern with it is the technical complexity of implementing API usage purposes, " & 26 & 12 & 14 \\ \addlinespace
5 & Other & & "I think it’s a nice improvement. I use AdMob, though, so I wonder how it would affect my advertising revenue." & 51 & 28 & 23 \\ \addlinespace
6 & No concern & Response blank or response does not include concerns about API usage purposes. & "None, looks declarative and straight forward" & 115 & 14 & 39 \\
\bottomrule
\end{tabularx}
\end{table*}

\clearpage

\clearpage
\section{Developer Survey}

\noindent This section presents the full list of questions and response options used in the developer survey. Data were collected via a Qualtrics survey. Response options were single-select form fields unless otherwise indicated. When the text shows "app/SDK", the survey actually shows either "app" or "SDK" depending on the response to Q3. 
\textcolor{black}{Horizontal lines indicate page breaks. To avoid answer order bias, response options were randomly ordered when appropriate and Likert scale options were randomly \textcolor{black}{reversed}.}

\vspace{-5pt}\hrulefill\par\vspace{0.10em}
\noindent
\textbf{Q0.} \textbf{[Consent.]}

\noindent
Your responses to this survey may be used in a research publication. All responses will be anonymized and any reports and presentations about the findings from this survey will not include any information that could identify you. 

\noindent
By continuing in this survey, you acknowledge you are at least 18 years of age and consent to the use of your responses in research publications.

\begin{itemize}\itemindent=-13pt
    \itemindent=-13pt
    \item [] $\square$ Continue
    \item [] $\square$ Exit
\end{itemize}

\vspace{-5pt}\hrulefill\par\vspace{0.1em}

\noindent
\textbf{Q1.} \textbf{[Multi-select with randomized response option order, excluding last option, "No".]}

\noindent
\textit{If the last option is selected, then the developer exits the survey.}

\noindent
Are you a software developer/engineer working on an Android application (app) or software development kit (SDK)?
    \begin{itemize}\itemindent=-13pt
    \itemindent=-13pt
    \item [] $\square$ Yes, I am a software developer/engineer working on an \textbf{Android app}
    \item [] $\square$ Yes, I am a software developer/engineer working on an \textbf{Android SDK}
    \item [] $\square$ No, I am not a software developer/engineer working on an Android app or SDK
\end{itemize}

\vspace{-5pt}\hrulefill\par\vspace{0.1em}

\noindent
\textbf{Q2.} \textbf{[Randomized response option order, excluding last option, "I don't know".]}

\noindent
\textit{This is a soft-screener question. Participants who answered incorrectly or "I don't know" were filtered out of the sample. The correct answer is "In the AndroidManifest.xml file using uses-permission tags".} 

\noindent
If an Android application needs access to potentially sensitive user data or system features (like location, camera, or contacts), where must the intent to use these permissions typically be declared?
        \begin{itemize}\itemindent=-13pt\itemindent=-13pt
    \itemindent=-13pt
    \item [] $\square$ Explicitly requested only in the Java/Kotlin code using a runtime permission check.
    \item [] $\square$ Within the project's .gitignore file to prevent accidental exposure.
    \item [] $\square$ In the application's build.gradle file, under defaultConfig.
    \item [] $\square$ In the AndroidManifest.xm file using uses-permission tags.
    \item [] $\square$ I don't know.
\end{itemize}

\vspace{-5pt}\hrulefill\par\vspace{0.1em}

\noindent
\textbf{Q3.} \textbf{[Randomized response option order.]}

\noindent
The following questions ask about the Android app or SDK you work on. If you work on multiple, please answer for the one you spend the most time on.

\noindent
Do you primarily work on an app, or on an SDK?
    \begin{itemize}\itemindent=-13pt
    \item [] $\square$ App
    \item [] $\square$ SDK
\end{itemize}

\vspace{-3pt}\hrulefill\par\vspace{0.1em}

\noindent
\textbf{Q4.}

\noindent
How many Android developers work on your App/SDK? If you work on more than one App/SDK, please answer for your primary App/SDK.
    \begin{itemize}\itemindent=-13pt
    \item [] $\square$ Just myself
    \item [] $\square$ 2-4
    \item [] $\square$ 5-9
    \item [] $\square$ 10-24
    \item [] $\square$ 25-49
    \item [] $\square$ 50 or more
\end{itemize}

\vspace{-3pt}\hrulefill\par\vspace{0.1em}

\noindent
\textbf{Q5.}\textbf{[Attention check.]}

\noindent
\textit{If the participant does not select "2 to 5" then they exit the survey.}

\noindent
This is an attention check. Select 2 to 5.
    \begin{itemize}\itemindent=-13pt
    \item [] $\square$ Just 1
    \item [] $\square$ 2 to 5
    \item [] $\square$ 6 to 10
    \item [] $\square$ More than 10
\end{itemize}

\vspace{-3pt}\hrulefill\par\vspace{0.1em}

\noindent
\textbf{Q6a.}\textbf{[Only shown if Q3 response was "App"; Dropdown using \href{https://support.google.com/googleplay/android-developer/answer/9859673}{list from Google Play store}]}

\noindent
Which of the following categories is your app listed as?

\hrulefill\par\vspace{0.1em}

\noindent
\textbf{Q6b.} \textbf{[Only shown if Q3 response was "SDK"; Options shown from \href{https://play.google.com/sdks}{Google Play SDK index categories}.]}

\noindent
Which of the following categories is your SDK listed as?

\hrulefill\par\vspace{0.1em}

\noindent
\textbf{Q7.}

\noindent
Which of the following applies to the app/SDK you work on?
    \begin{itemize}\itemindent=-13pt
    \item [] $\square$ Commercial / revenue generating
    \item [] $\square$ Open Source
    \item [] $\square$ Internal or enterprise use only
    \item [] $\square$ Other (please specify):
\end{itemize}

\vspace{-3pt}\hrulefill\par\vspace{0.1em}

\noindent
\textbf{Q8.}

\noindent
How many \textbf{years of professional experience} do you have as an Android developer?
    \begin{itemize}\itemindent=-13pt
    \item [] $\square$ Less than 1 year
    \item [] $\square$ 1 to 2 years
    \item [] $\square$ 3 to 5 years
    \item [] $\square$ 6 to 9 years
    \item [] $\square$ 10 to 14 years
    \item [] $\square$ 15+ years
\end{itemize}

\vspace{-3pt}\hrulefill\par\vspace{0.1em}

\noindent
\textbf{Q9.}

\noindent
What is your age in years?
    \begin{itemize}\itemindent=-13pt
    \item [] $\square$ 18-24
    \item [] $\square$ 25-34
    \item [] $\square$ 35-44
    \item [] $\square$ 45-54
    \item [] $\square$ 55-64
    \item [] $\square$ 65 or older
    \item [] $\square$ Prefer not to say
\end{itemize}

\hrulefill\par\vspace{4mm}

\noindent
\textbf{Q10.}

\noindent
What is your gender?
    \begin{itemize}\itemindent=-13pt
    \item [] $\square$ Woman
    \item [] $\square$ Man
    \item [] $\square$ Other
    \item [] $\square$ Prefer not to say
\end{itemize}

\hrulefill\par\vspace{0.1em}

\noindent
\textbf{Q11.} \textbf{[Optional; Drop down list of countries]}

\noindent
Please specify the country where you live.

\hrulefill\par\vspace{0.1em}

\noindent
\textbf{Q12.} \textbf{[Random ordering of the Android and Apple options. Likert scale options appeared horizontally in the survey.]}

\vspace{5pt}
\noindent
On a scale of 1 to 10 how much do you agree vs disagree with the following statements?



\begin{table}[h]
    \centering
    \small 
    \setlength{\tabcolsep}{1.5pt} 
    \begin{tabular}{|p{0.48\columnwidth}|p{0.48\columnwidth}|}
    \hline
    \textbf{Android} protects privacy & \textbf{Apple} protects privacy \\
    \hline
    1 - Strongly disagree & 1 - Strongly disagree \\
    \hline
    2 & 2 \\
    \hline
    3 & 3 \\
    \hline
    4 & 4 \\
    \hline
    5 & 5 \\
    \hline
    6 & 6 \\
    \hline
    7 & 7 \\
    \hline
    8 & 8 \\
    \hline
    9 & 9 \\
    \hline
    10 - Strongly agree & 10 - Strongly agree \\
    \hline
    \end{tabular}
\end{table}

\vspace{-2pt}
\hrulefill\par\vspace{0.1em}
\vspace{10pt}

\noindent
Device fingerprinting is a method to identify and track users. It involves:
\begin{enumerate}[label=(\alph*)]
    \item Collecting device-specific information from APIs
    \item Combining that information to create a unique "fingerprint" to identify the device
\end{enumerate}

\vspace{5pt}
\noindent
\textbf{Q13.}
\textbf{[Response order randomly \textcolor{black}{reversed}.]}

\noindent
Were you already familiar with device fingerprinting?
    \begin{itemize}\itemindent=-13pt
    \item [] $\square$ Very familiar
    \item [] $\square$ Somewhat familiar
    \item [] $\square$ I was not familiar with device fingerprinting, but now I understand it
    \item [] $\square$ Even after the explanation, I do not understand device fingerprinting
\end{itemize}

\hrulefill\par\vspace{0.1em}

\noindent
\textbf{Q14.} \textbf{[Response order randomly \textcolor{black}{reversed}.]}

\noindent
Does your app/SDK fingerprint users? \textit{Your answers are confidential. We will not attempt to connect your responses to you or your app/SDK.}

    \begin{itemize}\itemindent=-13pt
    \item [] $\square$ Yes
    \item [] $\square$ Not directly, but a dependency does
    \item [] $\square$ No
    \item [] $\square$ I'm not sure
\end{itemize}

\vspace{-3pt}\hrulefill\par\vspace{0.1em}

\noindent
\textbf{Q15.}

\noindent
The following questions explore a \textbf{hypothetical change to Android} called "API Usage Purposes". 

\noindent
This change is designed to protect users from unwanted fingerprinting, to improve user privacy. 

\noindent
With this change, Android developers would need to declare their purposes for specific APIs that can be used for fingerprinting. 

\noindent
The specific APIs would include the following.

\noindent
Does your app/SDK use these? Please mark all included in your app/SDK.

\begin{table}[h]
    \centering
    \scriptsize 
    \setlength{\tabcolsep}{2pt} 
    
    \resizebox{\columnwidth}{!}{%
        \begin{tabular}{|p{3.5cm}|c|c|c|c|}
        \hline
        \textbf{API} & \textbf{Yes} & \textbf{Dep.} & \textbf{No} & \textbf{Unsure} \\
        \hline
        {\color{DarkGreen}Settings.Secure.\allowbreak \color{DarkGreen} ANDROID\_ID} & $\square$ & $\square$ & $\square$ & $\square$ \\
        \hline
        {\color{DarkGreen}InputMethodManager.\allowbreak \color{DarkGreen} getEnabledInputMethodList} & $\square$ & $\square$ & $\square$ & $\square$ \\
        \hline
        {\color{DarkGreen}InputMethodManager.\allowbreak \color{DarkGreen} getEnabledInputMethodSubtypeList} & $\square$ & $\square$ & $\square$ & $\square$ \\
        \hline
        {\color{DarkGreen}InputMethodManager.\allowbreak \color{DarkGreen} getInputMethodList} & $\square$ & $\square$ & $\square$ & $\square$ \\
        \hline
        {\color{DarkGreen}TrafficStats.\allowbreak \color{DarkGreen} getTotalRxBytes} & $\square$ & $\square$ & $\square$ & $\square$ \\
        \hline
        {\color{DarkGreen}TrafficStats.\allowbreak \color{DarkGreen} getTotalRxPackets} & $\square$ & $\square$ & $\square$ & $\square$ \\
        \hline
        {\color{DarkGreen}TrafficStats.\allowbreak \color{DarkGreen} getTotalTxBytes} & $\square$ & $\square$ & $\square$ & $\square$ \\
        \hline
        {\color{DarkGreen}TrafficStats.\allowbreak \color{DarkGreen} getTotalTxPackets} & $\square$ & $\square$ & $\square$ & $\square$ \\
        \hline
        {\color{DarkGreen}AudioManager.\allowbreak \color{DarkGreen} getStreamVolume} & $\square$ & $\square$ & $\square$ & $\square$ \\
        \hline
        {\color{DarkGreen}AudioManager.\allowbreak \color{DarkGreen} getStreamMaxVolume} & $\square$ & $\square$ & $\square$ & $\square$ \\
        \hline
        {\color{DarkGreen}TelephonyManager.\allowbreak \color{DarkGreen} getSubscriberId} & $\square$ & $\square$ & $\square$ & $\square$ \\
        \hline
        {\color{DarkGreen}TelephonyManager.\allowbreak \color{DarkGreen} getNetworkCountryIso} & $\square$ & $\square$ & $\square$ & $\square$ \\
        \hline
        \end{tabular}%
    }
\end{table}

\vspace{-5mm}
\hrulefill\par\vspace{0.1em}
\noindent
\textit{API Usage Purposes} are designed to protect users from unwanted fingerprinting, to improve user privacy. 

\noindent
Android developers would add ``purposes'' to \textcolor{DarkGreen}{\texttt{<uses-permission>}} xml elements in their \textcolor{DarkGreen}{\texttt{AndroidManifest.xml}} file for specific APIs that can be used for fingerprinting.

\noindent
Here's an example of how API Usage Purposes would work. 

\noindent
Suppose you are the developer for an app/SDK that uses the APIs \textcolor{DarkGreen}{$\langle API\_1\rangle$} and \textcolor{DarkGreen}{$\langle API\_2\rangle$} from the previous table. 

\noindent
Assume there will be new permissions for each of these APIs. 

\noindent
Your \textcolor{DarkGreen}{\texttt{AndroidManifest.xml}} file would include the following:


{\color{DarkGreen}
\begin{Verbatim}[commandchars=\\\{\}, fontsize=\scriptsize]
<?xml version="1.0" encoding="utf-8"?>
<manifest xmlns:android="http://schemas.android.com/apk/res/android"
    xmlns:tools="http://schemas.android.com/tools"
    package="com.my.example.app">
    
    \textbf{<uses-permission}
            \textbf{android:name="android.permission.PERMISSION_FOR_API_1">}
        \textbf{<purpose android:name="NotForFingerprinting"/>}
    \textbf{</uses-permission>}
    
    \textbf{<uses-permission}
            \textbf{android:name="android.permission.PERMISSION_FOR_API_2">}
        \textbf{<purpose android:name="NotForFingerprinting" />}
    \textbf{</uses-permission>}
    ...
</manifest>
\end{Verbatim}
}

    
    



\noindent
Assume tools within Android Studio will help developers fill in the syntax for each API with automatic suggestions.

\vspace{10pt}
\noindent
\textbf{Q16.} \textbf{[Likert scale options appeared horizontally in the survey.]}

\noindent
\textbf{Suppose Android required API Usage Purposes} for APIs that can be used for fingerprinting. 

\noindent
Given this change, on a scale of 1 to 10 how much do you agree vs disagree with the below statement?



\begin{center}
\begin{tabular}{|p{0.5\columnwidth}|} 
\hline
\textbf{Android} protects user privacy \\
\hline
1 - Strongly disagree \hfill  \\
\hline
2 \hfill  \\
\hline
3 \hfill \\
\hline
4 \hfill  \\
\hline
5 \hfill  \\
\hline
6 \hfill  \\
\hline
7 \hfill  \\
\hline
8 \hfill \\
\hline
9 \hfill  \\
\hline
10 - Strongly agree \hfill  \\
\hline
\end{tabular}
\end{center}

\vspace{5pt}
\hrulefill\par\vspace{0.1em}
\vspace{5pt}

\noindent
\textit{[The order of the following two questions is randomized.]}

\vspace{5pt}
\noindent
\textbf{Q17a.} \textbf{[Optional; Open ended.]}

\noindent
What concerns, if any, do you have with API Usage Purposes?

\vspace{5pt}
\noindent
\textbf{Q17b.} \textbf{[Optional; Open ended.]}
\vspace{5pt}

\noindent
What benefits, if any, do you see with API Usage Purposes?

\hrulefill\par\vspace{0.1em}
\vspace{5pt}

\noindent
\textbf{Q18a-b.} \textbf{[The order of the following two questions is randomized; the order of response options is randomly \textcolor{black}{reversed}.]}

\noindent
\textbf{Suppose Android required API Usage Purposes} as described in this survey.

\noindent
What level of developer effort would be required to support API Usage Purposes?
    \begin{itemize}\itemindent=-13pt
    \item [] $\square$ Very little effort
    \item [] $\square$ Little effort
    \item [] $\square$ Moderate level of effort
    \item [] $\square$ Significant level of effort
    \item [] $\square$ Very significant level of effort\\
\end{itemize}

\noindent
What impact, if any, would this change have on user privacy?
    \begin{itemize}\itemindent=-13pt
    \item [] $\square$ It would have \textbf{a large negative impact} on user privacy
    \item [] $\square$ It would have \textbf{a small negative impact} on user privacy
    \item [] $\square$ It would have \textbf{no impact} on user privacy
    \item [] $\square$ It would have \textbf{a small positive impact} on user privacy
    \item [] $\square$ It would have \textbf{a large positive impact} on user privacy
\end{itemize}

\hrulefill\par\vspace{0.1em}

\noindent
\textbf{Q19.}

\noindent
Consider two ways Android could implement API Usage Purposes.

\noindent
\textbf{Optional model}: API Usage Purposes are optional.

\noindent
Apps that implement API Usage Purposes for all related APIs receive a privacy badge visible to users in the Google Play store and are ranked higher.

\noindent
\textbf{Required model}: API Usage Purposes are required.

\noindent
API calls will fail when API Usage Purposes are not provided for the impacted APIs.\\

\noindent
Do you think Android should implement API Usage Purposes?
    \begin{itemize}\itemindent=-13pt
    \item [] $\square$ Yes, with the \textbf{optional model}
    \item [] $\square$ Yes, with the \textbf{required model}
    \item [] $\square$ No, not at all
\end{itemize}

\hrulefill\par\vspace{0.1em}

\noindent
\textbf{Q20.} \textbf{[Optional; Open ended.]}

\noindent
Is there anything additional you'd like to share with Google about this hypothetical change (optional)?

\hrulefill\par\vspace{0.1em}

\noindent
\textbf{Q21.} \textbf{[Optional; Open ended.]}

\noindent
Do you have any comments on how to improve this survey (optional)?

\vspace{0.25cm}
\noindent Thank you for your time!

\end{document}